\documentclass{AIAA}
\usepackage{amsmath,amsthm}

\newtheorem{remark}{Remark}
\newtheorem{theorem}{Theorem}
\newtheorem{assumption}{Assumption}
\newtheorem*{pf}{Proof}

\begin{document}

\title{Design of distributed guidance laws for multi-UAV cooperative attacking a moving target based on reducing surrounding area
}
\author{Xiaoqian WEI\footnote{PhD Candidate, Science and Technology on Aerospace Intelligent Control Laboratory, State Key Laboratory for Turbulence and Complex Systems, Department of Mechanics and Engineering Science, College of Engineering, weixiaoqian@pku.edu.cn}, Jianying YANG\footnote{Professor, Science and Technology on Aerospace Intelligent Control Laboratory, State Key Laboratory for Turbulence and Complex Systems, Department of Mechanics and Engineering Science, College of Engineering, yjy@pku.edu.cn(corresponding author)} and Xiangru FAN\footnote{PhD, Clarkson University, Potsdam, Newyork, xiangrufan@gmail.com}}
\affiliation{Peking University, Beijing,  People's Republic of China, 100871}

\maketitle

\section{Abstract}


In this paper, two cooperative guidance laws based on the area around the target of multi-attackers are designed to deal with the problem of cooperative encirclement or simultaneous attack in the case of known target acceleration and unknown target acceleration. Multi-attacker communication network only needs to contain a directed spanning tree, and does not require all attackers to observe the target information, where at least one can observe the target. The components along the attacker-target line of sight in the novel guidance laws can reduce the relative remaining distance between the attacker and the target at the same speed, thus completing simultaneous attack and avoiding the calculation of the remaining time. The components of the guidance laws perpendicular to the attacker-target line of sight can make the normal overload of relative motion zero, so that the trajectory will not be distorted and the collision problem within the attacker group can be avoided. The simulation results verify the practicability of the novel guidance laws.

\section{Introduction}

 Distributed guidance laws can be applied to guide the multi-missile or multi-UAV system to encircle or attack a target simultaneously and to guide underwater multi-UAV system to capture a target cooperatively in \cite{1,2,3}. In these scenarios, target acceleration can be broadly categorized into 3 types, namely known acceleration in \cite{4,5,6,7,8,9,10,11,12}, unknown acceleration but with known upper bound in \cite{13,14,15}, and unknown acceleration but observable in \cite{16}. When the target acceleration is unknown but the upper bound is known, researchers usually utilize the adaptive coefficient method to cancel the disturbance caused by the unknown acceleration in \cite{14,15}. When the target acceleration is unknown but observable, the observer is designed to observe the target, so that the observation error can be small enough after a certain time in \cite{16}. In this paper, we primarily deal with two types of target acceleration mode: the known acceleration and the observable acceleration of the exogenous system.

At present, the most widely applied methods of cooperative attack are the remaining time coordination in \cite{17,18,19,20,21} and state (position and speed between individuals) coordination method in \cite{12,13,14}. One major drawback of remaining time coordination method is that the remaining time parameter is determined by the future state, which can not be accurately estimated since the target is maneuvering rapidly. As a consequence, the remaining time method can only deal with static or slow moving target with low acceleration in \cite{11,17,21}. The state coordination method, on the other hand, is unable to solve the problem of collision between attackers in the course of simultaneous encirclement due to the extreme spatial closeness between attackers. The guidance laws in this paper adopt the method of state coordination, and only use the relative motion information of the attackers and the target. Instead of requiring global information, the guidance laws we proposed only require a least one attacker to be able to observe the target, which greatly widens the applicability of novel guidance laws. The guidance laws are designed so that the enveloped area of multiple attackers can converge to the expected value, and they divide the relative motion of attacker-target into two sub-directions, each has
 their acceleration components play specific roles, which is rare in previous studies.


 In this paper, two sub-motion guidance laws are designed along the line-of-sight (LOS) direction and perpendicular to the LOS direction. The guidance law perpendicular to the LOS makes the angle of sight of the attacker change by controlling the acceleration component perpendicular to the LOS of the attacker-target relative motion, so that the area of the attacker surrounding the target asymptotically decrease until the expected area is reached, while the angular velocity of the LOS also converges to zero. The enclosure area is set to be less than the lethal area, and when it reaches zero, the multi-attacker cooperative attack on the target is completed. By controlling the acceleration component along the LOS of the attacker-target relative motion, the sub-motion guidance law along the LOS makes the relative motion remaining distance between the attacker-target consistent, and their relative velocities are the same negative value, so that the group movement of multiple attackers can satisfy the simultaneity.


The main contributions of this paper are as follows. Firstly, The collaborative guidance law is based on gradual reduction of enclosure area. Cooperative control is used to reduce enclosure area gradually, in which the coordination term is the relative distance between individual attackers and the target (the relative distance can always be obtained when the communication topology contains a spanning tree). As long as at least one attacker can obtain target information, other attackers can acquire the target information by information exchange between attackers and can calculate the relative distance between each attacker and target through geometric relationship between attackers. Therefore, this method can avoid the two major difficulties of previous methods: calculation of residual time and avoidance of collision.


Secondly, the control input of the guidance law can be divided into the normal and tangential acceleration along the LOS. Saturation attack can be guaranteed if the area of the attacker around the target is reduced to zero, which can be achieved by using solely the normal direction control term. In additional, that the normal acceleration decreases to zero can ensure the trajectory smoothness and can enhance the applicability of the guidance law. On this basis, the tangential acceleration control is added to control the instantaneous relative distances and relative velocities between attackers and the target to ensure simultaneous hit. Therefore, this method guarantees simultaneous hit in addition to saturation attack.


Thirdly, the novel distributed adaptive cooperative guidance law can function properly even when information communication is restricted to be only between neighboring nodes. The guidance law does not require the acquisition of characteristics information of the whole communication topology. Therefore, it can adapt to the arbitrary changes of the communication topology and automatically adjust the communication topology connection structure with the change of the distance between each individual, or even adapt to the communication. The only guarantee for the intermittent connectivity of the communication topology is that the communication network contains a spanning tree, which greatly reduces the requirement of the fixed topology commonly used in cooperative guidance methods in the past.

 The rest of this paper is organized as follows. The problem statement is given in the next section. Section \uppercase\expandafter{\romannumeral3} and Section \uppercase\expandafter{\romannumeral4} present the analyses of design of distributed guidance laws for multi-UAV cooperative attacking a moving target with known acceleration and unknown acceleration respectively based on reducing surrounding area. Numerical simulation is shown in Section \uppercase\expandafter{\romannumeral5}, the main contributions of the paper are summarized in Section \uppercase\expandafter{\romannumeral6}.

\vspace{-6pt}

\section{Preliminaries}
In the field of multi-agent system, the communication topology of a multi-agent system is described as a directed graph. Individual agent is treated as a vertex in the communication graph and the information communication link between two adjacent agents is modelled as an edge in the communication graph.

Take a multi-agent network consisting of N agents as an example. The directed graph $\textbf{\emph{G}}(\textbf{\emph{V}},\textbf{\emph{E}},\textbf{\emph{A}})$ is used to describe the communication topology, $\textbf{\emph{E}}\subseteq \textbf{\emph{V}}\times\textbf{\emph{V}}$ is the set of edges and the matrix $\textbf{\emph{A}}=[a_{ij}]_{N\times N}$ with elements $a_{ij}$ is the weighted adjacency matrix. An edge represents an information link between an ordered pair of nodes $(i,j)$, which stands for node j to node i, in the communication topology $\textbf{\emph{G}}$. Self-loops are not permitted in the communication topology, which means $(i,i)$ is not allowed. A directed route from node i to node j is defined as a sequence of paths, $(i,k_1)$,$(k_1,k_2)$,...,$(k_l,j)$, with different nodes $k_m, m=1,2,...,l$.

 A graph is called undirected if and only if there exists an edge $(j,i)$ in $\textbf{\emph{E}}$ for any $(i,j)\in\textbf{\emph{E}}$. This structure is equivalent to a spanning tree, which is a directed rooted tree that utilizes a directed path starting from the root vertex to connect every other vertex in the graph. When there are undirected paths between any pair of different vertices in undirected graphs, undirected graphs are considered to be connected; similarly, in directed graphs, directed graphs are considered to be strongly connected. Strongly connected graphs must contain a spanning tree, but not vice versa. Moreover, Laplacian matrix $\textbf{\emph{L}}=[l_{ij}]_{N\times N}$ of $\textbf{\emph{G}}$ concerning adjacency matrix $\textbf{\emph{A}}$ is characterized as $l_{ij}=-a_{ij}, i\neq j$ and $l_{ii}=\sum_{j=1,j\neq i}^Na_{ij}$.

\section{Problem formulation}
\vspace{-2pt}
\begin{figure}[!htb]
 \centering
 \includegraphics[width=0.5\textwidth]{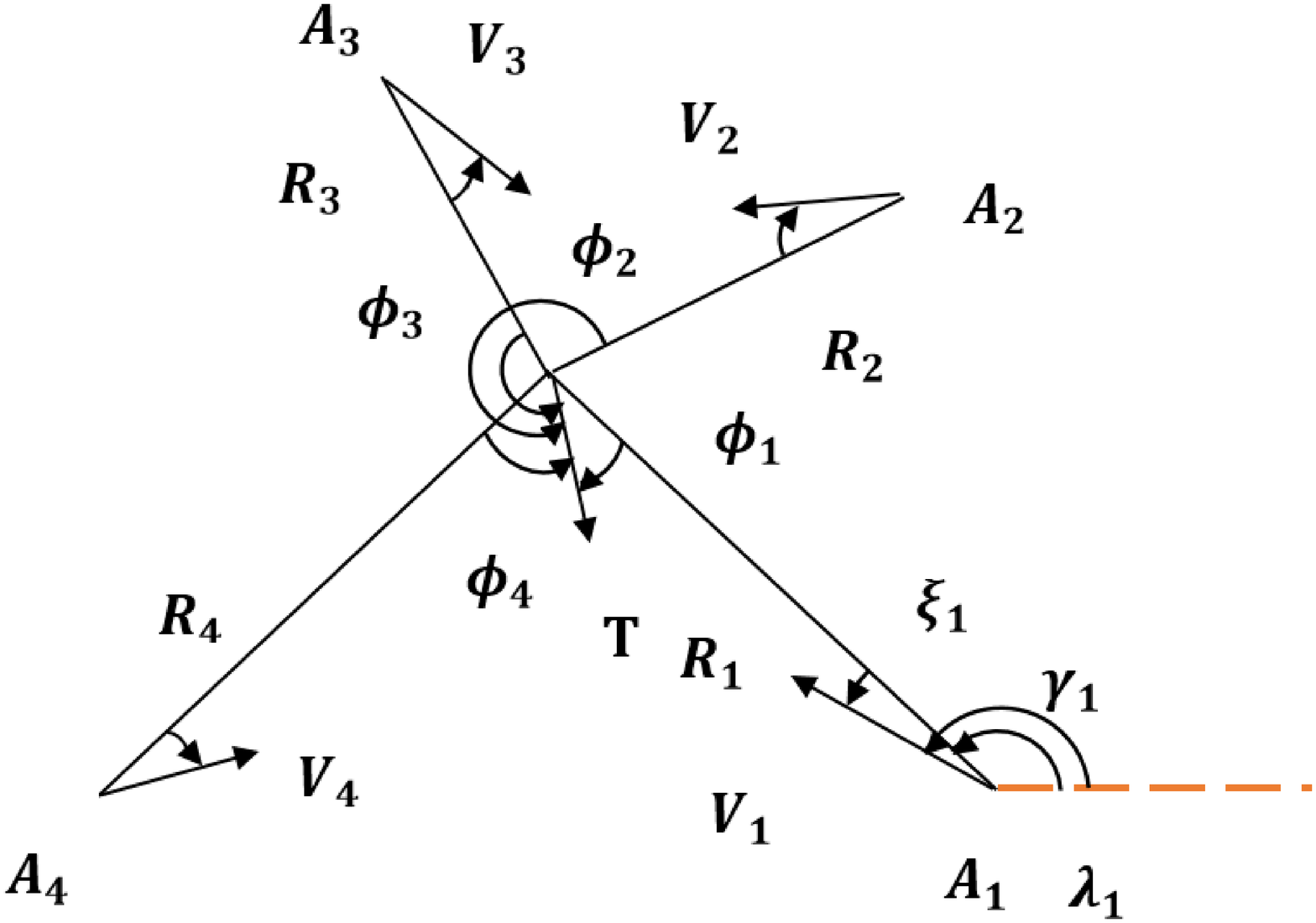}
 \caption{Geometry for TA engagement.}
 \label{Fig1}
\end{figure}

\begin{figure}[!htb]
 \centering
 \includegraphics[width=0.4\textwidth]{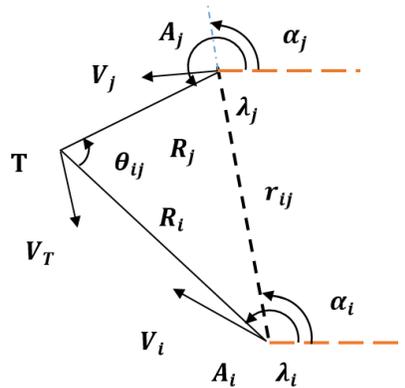}
 \caption{Geometry for an attacker and its neighbors.}
 \label{Fig2}
\end{figure}

\begin{figure}[!htb]
 \centering
 \includegraphics[width=0.45\textwidth]{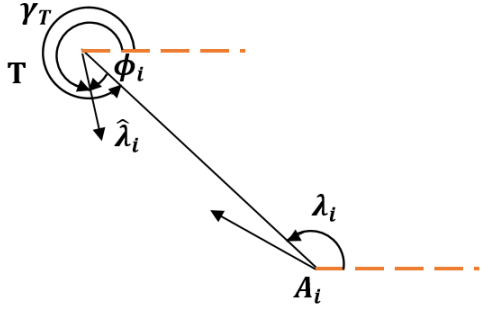}
 \caption{The relationship among angles $\delta_T$, $\phi_i$ and ${\hat{{\lambda}}_i}$.}
 \label{Fig3}
\end{figure}


In this section, we consider the TA scenario where multi-attackers intercept one maneuvering target in two-dimensional space in Fig.1. Multi-attackers with constant speed are denoted as nodes ${\mathcal{V}}={1,...,N}$. For simplicity, we show only a geometry for the i-th attacker and its neighbors in Fig.2. The relationship between the angles $\delta_T$, $\phi_i$ and ${\hat{{\lambda}}_i}$ is described in Figure 3. Attention should be paid to the fact that the acceleration direction of the attacker and the target is always perpendicular to their respective velocity direction, that is, their velocity values are determined ahead of time and the directions are time-varying. In this paper, the speed of the attacker is always smaller than that of the target.

The dynamic equations of the i-th attacker are given by
\begin{equation}\label{eq:1}
\begin{split}
&\dot{R_i}=V_{ri}, V_{ri}=V_T{\cos}{\phi}_{i}-V_i{\cos}{\xi}_{i}\\
&\dot{{\lambda}_i}=\frac{V_{\lambda i}}{R_i}, V_{\lambda i}=V_T{\sin}{\phi}_{i}-V_i{\sin}{\xi}_{i}\\
&{\gamma}_{i}={\xi_i}+{{\lambda}_i}, \gamma_T=\phi_i+\hat{{\lambda}}_i\\
&\dot{{\gamma}_{i}}=\frac{A_{Mi}}{V_i}, \dot{{\gamma}_{T}}=\frac{A_{T}}{V_T}\\
&{\theta_{ij}}={\lambda}_j-{\lambda}_i
\end{split}
\end{equation}
with
\begin{center}
${\hat{{\lambda}}_i}$=$\left[\begin{array}{lll} {\lambda}_i-\pi & if & {\lambda}_i\geq \pi\\ {\lambda}_i+\pi & if & {\lambda}_i< \pi\end{array}\right]$
\end{center}
where the subscripts i and T denote the i-th attacker and the target. ${V_{ri}}$ are the attacker-target relative velocity components along the LOS, and ${V_{\lambda i}}$ are the attacker-target relative velocity components normal to the LOS. ${{\lambda}_i}$ is the LOS angle of the i-th attacker in the inertial reference frame, and ${\theta_{ii+1}}$ is the angle from the i-th LOS to the i+1-th LOS. The terms ${\sigma}_{i}$ and ${\sigma}_{T}$ are the heading angles of the attacker and the target. $R_i$ is the relative distance between the i-th attacker and the target, and $r_{ii+1}$ is the relative distance between the i-th and the i+1-th attacker. ${\xi}_{i}$ is the bearing angle between the i-th LOS and the direction of the i-th attacker's velocity, while the term ${\phi_i}$ is the bearing angle between the i-th LOS and the direction of the target's velocity. Positive constants $V_i$ and $V_T$ are the velocities of the i-th attacker and the target. The equations of motion for other attackers are similar.

The derivative of the above formulas are
\begin{equation}\label{eq:2}
\begin{split}
&\dot{V_{ri}}=\frac{{V^2_{\lambda i}}}{{R_i}}-{A_{Mri}}+{A_{Tri}}\\
&\dot{V_{\lambda i}}=-\frac{{V}_{\lambda i}{V}_{ri}}{R_i}-{A}_{M{\lambda}i}+{A}_{T{\lambda}i}
\end{split}
\end{equation}
where ${A_{Tri}}=-A_T\sin\phi_i$ are the target acceleration components along the LOS, while ${A_{T\lambda i}}=A_T\cos\phi_i$ are the target acceleration components normal to the LOS. ${A_{Mri}}=A_{Mi}\sin\xi_i$ are the i-th attacker's acceleration components along the LOS, while ${A_{M\lambda i}}=-A_{Mi}\cos\xi_i$ are the i-th attacker's acceleration components normal to the LOS.

\begin{assumption}\label{the1}{\rm
The graph $\textbf{\emph{G}}$ that describes the communication topology of the multi-agents system in this paper is directed and contains a spanning tree.
}
\end{assumption}

\begin{assumption}\label{the2}{\rm
At least one attacker can observe the relative distance, relative azimuth and relative velocity between the attacker and the target.
}
\end{assumption}

 \begin{remark}
\label{rem1}
{\rm

If the multi-attacker group satisfies Assumption 2, that is, the target velocity $V_T$, relative position $R_i$ and LOS angle $\lambda_i$ can be observed by i-th attacker, then the neighbor attacker j of i-th attacker in the detection range (i.e. $j \in \boldsymbol{N}_i$) can also obtain information $V_T$, $R_i$, $\lambda_i$, the distance between itself and i-th attacker $r_{ij}$ and angle $\alpha_j$ (Note that $\alpha_i=\alpha_j$) is shown in Figure 2. Then, j-th attacker can get the relative distance $R_j=\sqrt{R_i^2+r_{ij}^2-2R_ir_{ij}\cos(\lambda_i-\alpha_i)}$ and LOS angle $\lambda_j=\lambda_i+\arcsin(\frac{r_{ij}\sin(\lambda_i-\alpha_i)}{R_j})$ according to triangle sine theorem and triangle cosine theorem. Similarly, the neighbor attacker k of j-th attacker in the detection range (i.e. $k \in \boldsymbol{N}_j$) can also get the relative distance $R_k$ and LOS angle $\lambda_k$. According to Assumption 1, all attackers can get relative distance and relative speed between themselves and the target.
}
 \end{remark}

\section{Analysis}


In this section, we discuss how low-speed attackers can cooperatively enclose or hit a high-speed target under conditions of both known and unknown target acceleration, and design two guidance laws based on enclosing area to attack or surround the target simultaneously. The problem of UAVs attacking a moving target simultaneously is solved by using the acceleration of target-attacker relative motion. The relative movement of the target-attacker is divided into two sub-directions: normal to the LOS and tangential to the LOS. The acceleration component normal to the LOS can make the enclosure area of the target surrounded by multiple attackers converge to the expected value, and can make the angular velocity of the LOS angle zero, which can avoid the early collision caused by changing the LOS angle in the process of attack. The acceleration tangential to LOS can force the relative distance between multiple attackers and the target to be the same when they successfully surrounded the target. And it can also control the relative speed between attackers and target to reach the same negative value, that is, the relative distance decreases at the same speed uniformly, so that multiple attackers can attack the target at the same time. Note that relative distance $R_i$, $R_j$, attackers' velocity $V_i$, $V_j$, and target's velocity $V_T$ are positive real numbers by definition.

\subsection{Cooperative attack with known target acceleration }
In this section, we discuss how multiple attackers can encircle or attack a maneuvering target with known acceleration simultaneously.

Since the attacker's sensor can usually obtain the relative distance $R$, relative velocity $V_r$, $V_\lambda$ and relative LOS angle $\lambda$ of the target, and the position and velocity of the attacker itself are known, the attacker can calculate the position and velocity of the target timely, and then calculate the acceleration of the target. Although using estimations will inevitably introduce errors and delays, they can be compensated by the observer. Therefore, we assume that the target acceleration information is known in this theoretical study, which is consistent with most application scenarios. In next subsection, we will further consider how to estimate the acceleration by the relative distance, relative velocity and relative LOS angle of the target.

\begin{assumption}\label{the13}{\rm
The acceleration of the target is known.
}
\end{assumption}

\begin{theorem}\label{the1}
{\rm
For the problem of multiple attackers encircling or attacking a moving target with known acceleration at the same time in formulas~(\ref{eq:1},\ref{eq:2}) and the information transmission network satisfies Assumption 1, the acceleration of normal and tangential relative motion and the corresponding adaptive parameters can be designed as follows

\begin{equation}\label{eq:3}
\begin{split}
&{A_{M\lambda i}}=-\frac{{V}_{ri}{V}_{\lambda i}}{R_i}+{{A}_{T{\lambda}i}}\\
&+\kappa_{1}(1+\sum_{j=1}^N\frac{a_{ij}(R_iR_j\sin\theta_{ij})^2}{R_{i0}^2R_{j0}^2})^{-1}\sum_{j=1}^N \frac{a_{ij}V_{\lambda i}}{R_{i0}^2R_{j0}^2}(V_TR_iR_j^2+V_TR_i^2R_j+V_iR_iR_j^2+V_jR_i^2R_j)\\
&{A_{Mri}}=\frac{{V^2_{\lambda i}}}{{R_i}}+{A_{Tri}}+\kappa_2(V_{ri}+c)+R_i\\
&\dot{\mu_i}=(1+(V_{ri}+c)^2)^{-1}\mu_iR_i(V_{ri}+c)
\end{split}
\end{equation}
where $\kappa_1>1$, $\kappa_2>0$, and $c\geq0$.
}
\end{theorem}

 \begin{pf}{\rm
In the normal direction of the LOS, we use the following Lyapunov function $V_1(t)$
\begin{equation}\label{eq:4}
\begin{split}
&V_1(t)=\frac{1}{2}\sum_{i=1}^N\sum_{j=1}^N\frac{a_{ij}V_{\lambda i}^2}{R_{i0}^2R_{j0}^2}(R_iR_j\sin\theta_{ij})^2+\frac{1}{2}\sum_{i=1}^NV_{\lambda i}^2
\end{split}
\end{equation}

Then, the derivative of $V_1(t)$ along the direction of equation~(\ref{eq:3}) is shown as
\begin{equation}\label{eq:5}
\begin{split}
&\dot{V_1}(t)=\sum_{i=1}^N\sum_{j=1}^N\frac{a_{ij}V_{\lambda i}^2}{R_{i0}^2R_{j0}^2}(R_iR_j\sin\theta_{ij})(R_iR_j\sin\theta_{ij})^{'}+\sum_{i=1}^N(1+\sum_{j=1}^N\frac{a_{ij}(R_iR_j\sin\theta_{ij})^2}{R_{i0}^2R_{j0}^2})V_{\lambda i}\dot{V_{\lambda i}}
\end{split}
\end{equation}

By substituting equation~(\ref{eq:3}) into equation~(\ref{eq:2}), we can get
\begin{equation}\label{eq:6}
\begin{split}
&\dot{V_{\lambda i}}=-\kappa_{1}(1+\sum_{j=1}^N\frac{a_{ij}(R_iR_j\sin\theta_{ij})^2}{R_{i0}^2R_{j0}^2})^{-1}\sum_{j=1}^N \frac{a_{ij}V_{\lambda i}}{R_{i0}^2R_{j0}^2}(V_TR_iR_j^2+V_TR_i^2R_j+V_iR_iR_j^2+V_jR_i^2R_j)\\
&\dot{V_{ri}}=-\kappa_2(V_{ri}+c)-R_i
\end{split}
\end{equation}

Note that
\begin{equation}\label{eq:7}
\begin{split}
&(R_iR_j\sin\theta_{ij})^{'}=(\dot{R_i}R_j+R_i\dot{R_j})\sin\theta_{ij}+R_iR_j\cos\theta_{ij}\dot{\theta_{ij}}\\
&=(V_T{\cos}{\phi}_{i}-V_i{\cos}{\xi}_{i})R_j\sin\theta_{ij}+(V_T{\cos}{\phi}_{j}-V_j{\cos}{\xi}_{j})R_i\sin\theta_{ij}\\
&+(V_T{\sin}{\phi}_{j}-V_j{\sin}{\xi}_{j})R_i\cos\theta_{ij}-(V_T{\sin}{\phi}_{i}-V_i{\sin}{\xi}_{i})R_j\cos\theta_{ij}\\
&=(R_j{\cos}{\phi}_{i}+R_i{\cos}{\phi}_{j})V_T\sin\theta_{ij}-(V_iR_j{\cos}{\xi}_{i}+V_jR_i{\cos}{\xi}_{j})\sin\theta_{ij}\\
&-(V_T{\sin}{\phi}_{i}-V_i{\sin}{\xi}_{i})R_j\cos\theta_{ij}+(V_T{\sin}{\phi}_{j}-V_j{\sin}{\xi}_{j})R_i\cos\theta_{ij}\\
&=R_jV_T(\sin\theta_{ij}{\cos}{\phi}_{i}-\cos\theta_{ij}{\sin}{\phi}_{i})+R_iV_T(\sin\theta_{ij}{\cos}{\phi}_{j}+\cos\theta_{ij}{\sin}{\phi}_{j})\\
&-V_iR_j({\cos}{\xi}_{i}\sin\theta_{ij}-{\sin}{\xi}_{i}\cos\theta_{ij})-V_jR_i({\sin}{\xi}_{j}\cos\theta_{ij}+{\cos}{\xi}_{j}\sin\theta_{ij})\\
&=R_jV_T\sin(\theta_{ij}-{\phi}_{i})+R_iV_T\sin(\theta_{ij}+{\phi}_{j})-V_iR_j\sin(\theta_{ij}-{\xi}_{i})-V_jR_i\sin(\theta_{ij}+{\xi}_{j})\\
&=R_jV_T\sin\beta_{1i}+R_iV_T\sin\beta_{2i}-V_iR_j\sin\beta_{3i}-V_jR_i\sin\beta_{4i}
\end{split}
\end{equation}
where $\beta_{1i}=\theta_{ij}-{\phi}_{i}$, $\beta_{2i}=\theta_{ij}+{\phi}_{j}$, $\beta_{3i}=\theta_{ij}-{\xi}_{i}$, and $\beta_{4i}=\theta_{ij}+{\xi}_{j}$.

From equations~(\ref{eq:6},\ref{eq:7}), we can obtain that
\begin{equation}\label{eq:8}
\begin{split}
&\dot{V_1}(t)=-\sum_{i=1}^N\sum_{j=1}^N\frac{a_{ij}V_{\lambda i}^2}{R_{i0}^2R_{j0}^2}[V_TR_iR_j^2(\kappa_1-\sin\theta_{ij}\sin\beta_{1i})+V_TR_i^2R_j(\kappa_1-\sin\theta_{ij}\sin\beta_{2i})\\
&+V_iR_iR_j^2(\kappa_1+\sin\theta_{ij}\sin\beta_{3i})+V_jR_i^2R_j(\kappa_1+\sin\theta_{ij}\sin\beta_{4i}))]<0
\end{split}
\end{equation}
with $\kappa_1>1$. Note that relative distance $R_i$, $R_j$, attackers' velocity $V_i$, $V_j$, and target's velocity $V_T$ are positive when defined.

Therefore, using the normal acceleration $A_{M \lambda}$ in formula~(\ref{eq:3}), the area of multiple attackers surrounding the target $S=\frac{1}{4}\sum_{i=1}^N\sum_{j=1}^Na_{ij}|R_iR_j\sin\theta_{ij}|$ and the LOS angular velocity $\dot{\lambda}=\frac{V_\lambda}{R}$ of multiple attackers will converge to zero.

In the tangential direction of the LOS, we use the following Lyapunov function $V_2(t)$
\begin{equation}\label{eq:9}
\begin{split}
&V_2(t)=\frac{1}{2}\sum_{i=1}^N\sum_{j=1}^N a_{ij}[(R_i-R_j)^2+(V_{ri}-V_{rj})^2]+\frac{1}{2}\sum_{i=1}^N\mu_i^2(V_{ri}+c)^2+\frac{1}{2}\sum_{i=1}^N\mu_i^2
\end{split}
\end{equation}

Then, the derivative of $V_2(t)$ along the direction of equation~(\ref{eq:6}) is shown as
\begin{equation}\label{eq:10}
\begin{split}
&\dot{V_2}(t)=\sum_{i=1}^N\sum_{j=1}^N a_{ij}[(R_i-R_j)(V_{ri}-V_{rj})+(V_{ri}-V_{rj})(\dot{V_{ri}}-\dot{V_{rj}})]+\sum_{i=1}^N\mu_i^2(V_{ri}+c)\dot{V_{ri}}+\sum_{i=1}^N(1+(V_{ri}+c)^2)\mu_i\dot{\mu_i}\\
&=-\kappa_2\sum_{i=1}^N\sum_{j=1}^N a_{ij}(V_{ri}-V_{rj})^2-\kappa_2\sum_{i=1}^N\mu_i^2(V_{ri}+c)^2<0
\end{split}
\end{equation}
with $\kappa_2>0$.

Therefore, using the tangential acceleration $A_{Mr}$ in formula~(\ref{eq:3}), we can obtain $R_i=R_j$ and $V_{ri}=-c, c\geq0$, which means multiple attackers can attack the target at the same time at relatively uniform speed $V_r=-c$.
$\qed$
 }
 \end{pf}

\begin{remark}
\label{rem2}
{\rm
The guidance law in this paper is the acceleration of relative motion between multiple attackers and a target. Its essence is a improved proportional guidance, that is, the first item of ${A_{M\lambda i}}$ in formula~(\ref{eq:3}) is proportional to $\dot{\lambda_i}$ and the ratio is $-{V}_{ri}$, and the other two items can be regarded as correction items.
}
 \end{remark}

  \begin{remark}
\label{rem3}
{\rm

If only the acceleration component perpendicular to the LOS $A_{M \lambda}$ is controlled, that is, changing the attacker's LOS angle $\lambda$ can make the envelope area of the attacker $S=\frac{1}{4}\sum_{i=1}^N\sum_{j=1}^Na_{ij}|R_iR_j\sin\theta_{ij}|$ reach the expected value $S_c=\pi R_c^2$ with the killing radius $R_c$ (we take $R_c=0$) and the LOS angular velocity $\dot{\lambda}=\frac{V_\lambda}{R}$ converges to zero, it can also make multiple attackers encircle or attack the target cooperatively. It should be noted that $R_{i0}^2R_{j0}^2$ on the denominator of formula~(\ref{eq:4}) is to avoid the excessive value of the area term $\frac{1}{2}\sum_{i=1}^N\sum_{j=1}^N\frac{a_{ij}V_{\lambda i}^2}{R_{i0}^2R_{j0}^2}(R_iR_j\sin\theta_{ij})^2$, which leads to the excessive gain of the guidance law $A_{M \lambda}$. At this time, the LOS angular velocity is zero (i.e., the normal overload $\dot{V_\lambda}$ is zero), which means that the LOS angle of each attacker will not change significantly before completing the encirclement or attack task, and the trajectory is smooth, thus effectively avoiding the internal collision problem of multiple attackers in advance. It is important to note that multiple attackers do not necessarily hit the target at the same time, or the enclosure area is not circular.


If only the acceleration component along the LOS $A_{Mr}$ is controlled, that is, the remaining distance $R$ between the attacker and the target along the LOS is consistent (i.e., $R_i=R_j, j\in \boldsymbol {N}_i$), and the remaining distance of all attackers decreases at the same speed $V_{ri}=V_{rj}=-c$ with $c\geq0$, so that multiple attackers can encircle (the encircling area is circular) or attack the target simultaneously. It should be noted that the normal overload $\dot{V_\lambda}$ is not necessarily zero at this time, multiple attackers are likely to collide in advance, and the trajectory is not necessarily smooth and may not have practical significance.

To sum up, both the tangential and normal acceleration components of the attacker-target relative movement along LOS ought to be controlled, the task of multiple attackers that encircling or simultaneously attacking a moving target will be fulfilled.
}
 \end{remark}

 \subsection{Cooperative attack with unknown target acceleration}
In this section, we discuss how multiple attackers can encircle or attack a maneuvering target with unknown acceleration at the same time. The target acceleration is unknown, but the change structure of the target acceleration is known, and its initial condition is unknown. Therefore, the current acceleration of the target is unknown. This is a common method to deal with an unknown target, which has certain practical significance. Although the maneuvering information of the actual target is unknown, the type of the target (such as aircraft or motor vehicles) is known. Therefore, we can assume that the basic characteristic structure of its motion is known. This approach is also in line with most of the actual situations.


 Assuming that
\begin{equation}\label{eq:11}
\begin{split}
&\dot{A_T}=sA_T
\end{split}
\end{equation}
which means it is an exogenous system, where $s\leq0$ is a known constant. It ought to be noticed that while $s$ is greater than zero, namely, the target's acceleration does not converge to an upper bound. This situation rarely occurs in practice since it does not satisfy the physical boundaries.

Distributed disturbance observer is designed in the following form

\begin{equation}\label{eq:12}
\begin{split}
&\dot{z_i}=sz_i+\sigma_{1i}V_{\lambda i}+\sigma_{2i}V_{ri}\\
&z_i=\hat{A_{Ti}}
\end{split}
\end{equation}
where $\hat{A_{Ti}}$ is the target acceleration observed by i-th attacker. $\sigma_{1i}$ and $\sigma_{2i}, i \in \textbf{\emph{V}}={1,2,...,N}$ are the observer coefficients to be determined later, and $z_i$ is the virtual state of the disturbance observer.

The following shows that selecting the right $\sigma_{1i}$ and $\sigma_{2i}, i \in \textbf{\emph{V}}={1,2,...,N}$ allows the realization of consistency of $V_{\lambda i}$, $V_{ri}$ and $R_i$, $\forall i, j \in \textbf{\emph{V}}={1,2,...,N}$.

\begin{theorem}\label{the2}
{\rm
For the problem of multiple attackers encircling or attacking a moving target whose acceleration is in the form of equation~(\ref{eq:11}) simultaneously in formulas~(\ref{eq:1},\ref{eq:2}) and the information transmission network satisfies Assumption 1, the acceleration of normal and tangential relative motion can be designed in equation~(\ref{eq:13}) along with the distributed disturbance observer in equation~(\ref{eq:12}) whose observer coefficients satisfy $\sigma_{1i}=(1+\sum_{j=1}^N\frac{a_{ij}(R_iR_j\sin\theta_{ij})^2}{R_{i0}^2R_{j0}^2})\cos{\phi}_{i}$ and $\sigma_{2i}=-\sin{\phi}_{i}, i \in \textbf{\emph{V}}={1,2,...,N}$.

\begin{equation}\label{eq:13}
\begin{split}
&{A_{M\lambda i}}=-\frac{{V}_{ri}{V}_{\lambda i}}{R_i}+\hat{A_{T\lambda i}}\\
&+\kappa_{1}(1+\sum_{j=1}^N\frac{a_{ij}(R_iR_j\sin\theta_{ij})^2}{R_{i0}^2R_{j0}^2})^{-1}\sum_{j=1}^N \frac{a_{ij}V_{\lambda i}}{R_{i0}^2R_{j0}^2}(V_TR_iR_j^2+V_TR_i^2R_j+V_iR_iR_j^2+V_jR_i^2R_j)\\
&{A_{Mri}}=\frac{{V^2_{\lambda i}}}{{R_i}}+\hat{A_{Tri}}+\kappa_2V_{ri}+R_i\\
\end{split}
\end{equation}
where $\kappa_1>1$, $\kappa_2>0$, $c\geq0$, $\hat{A_{T\lambda i}}=\hat{A_{Ti}}\cos\phi_i$ and $\hat{A_{Tri}}=-\hat{A_{Ti}}\sin\phi_i$.
}
\end{theorem}

 \begin{pf}{\rm
We use the following Lyapunov function $V_3(t)$
\begin{equation}\label{eq:14}
\begin{split}
&V_3(t)=\frac{1}{2}\sum_{i=1}^N\sum_{j=1}^N\frac{a_{ij}V_{\lambda i}^2}{R_{i0}^2R_{j0}^2}(R_iR_j\sin\theta_{ij})^2+\frac{1}{2}\sum_{i=1}^NV_{\lambda i}^2+W(t)\\
&W(t)=\frac{1}{2}\sum_{i=1}^NR_i^2+\frac{1}{2}\sum_{i=1}^NV_{ri}^2+\frac{1}{2}\sum_{i=1}^N\tilde{A_{Ti}}^2
\end{split}
\end{equation}
with $\tilde{A_{Ti}}=A_T-\hat{A_{Ti}}$.

By substituting equation~(\ref{eq:13}) into equation~(\ref{eq:2}), we can get
\begin{equation}\label{eq:15}
\begin{split}
&\dot{V_{\lambda i}}=\tilde{A_{T\lambda i}}-\kappa_{1}(1+\sum_{j=1}^N\frac{a_{ij}(R_iR_j\sin\theta_{ij})^2}{R_{i0}^2R_{j0}^2})^{-1}\sum_{j=1}^N \frac{a_{ij}V_{\lambda i}}{R_{i0}^2R_{j0}^2}(V_TR_iR_j^2+V_TR_i^2R_j+V_iR_iR_j^2+V_jR_i^2R_j)\\
&\dot{V_{ri}}=\tilde{A_{Tri}}-\kappa_2V_{ri}-R_i
\end{split}
\end{equation}
with $\tilde{A_{T\lambda i}}=A_{T\lambda i}-\hat{A_{T\lambda i}}=\tilde{A_{Ti}}\cos\phi$ and $\tilde{A_{Tri}}=A_{Tri}-\hat{A_{Tri}}=-\tilde{A_{Ti}}\sin\phi$.

Then, the derivative of $V_3(t)$ along the direction of equation~(\ref{eq:15}) is shown as
\begin{equation}\label{eq:16}
\begin{split}
&\dot{V_3}(t)=\sum_{i=1}^N\sum_{j=1}^N\frac{a_{ij}V_{\lambda i}^2}{R_{i0}^2R_{j0}^2}(R_iR_j\sin\theta_{ij})(R_iR_j\sin\theta_{ij})^{'}+\sum_{i=1}^N(1+\sum_{j=1}^N\frac{a_{ij}(R_iR_j\sin\theta_{ij})^2}{R_{i0}^2R_{j0}^2})V_{\lambda i}\dot{V_{\lambda i}}+\dot{W(t)}\\
&=-\sum_{i=1}^N\sum_{j=1}^N\frac{a_{ij}V_{\lambda i}^2}{R_{i0}^2R_{j0}^2}[V_TR_iR_j^2(\kappa_1-\sin\theta_{ij}\sin\beta_{1i})+V_TR_i^2R_j(\kappa_1-\sin\theta_{ij}\sin\beta_{2i})\\
&+V_iR_iR_j^2(\kappa_1+\sin\theta_{ij}\sin\beta_{3i})+V_jR_i^2R_j(\kappa_1+\sin\theta_{ij}\sin\beta_{4i}))]+\sum_{i=1}^N(1+\sum_{j=1}^N\frac{a_{ij}(R_iR_j\sin\theta_{ij})^2}{R_{i0}^2R_{j0}^2})V_{\lambda i}\tilde{A_{T\lambda i}}+\dot{W(t)}\\
&\leq\sum_{i=1}^N(1+\sum_{j=1}^N\frac{a_{ij}(R_iR_j\sin\theta_{ij})^2}{R_{i0}^2R_{j0}^2})V_{\lambda i}\tilde{A_{T\lambda i}}+\dot{W(t)}
\end{split}
\end{equation}

Note that
\begin{equation}\label{eq:17}
\begin{split}
&\dot{W(t)}=\sum_{i=1}^NR_iV_{ri}+\sum_{i=1}^NV_{ri}\dot{V_{ri}}+\sum_{i=1}^N\tilde{A_{Ti}}\dot{\tilde{A_{Ti}}}\\
&=-\kappa_2\sum_{i=1}^NV_{ri}^2+s\sum_{i=1}^N\tilde{A_{Ti}}^2-\sum_{i=1}^N\tilde{A_{Ti}}V_{ri}(\sin\phi_i+\sigma_{2i})-\sum_{i=1}^N\tilde{A_{Ti}}V_{\lambda i}\sigma_{1i}
\end{split}
\end{equation}

From equations~(\ref{eq:16},\ref{eq:17}), we can obtain that
\begin{equation}\label{eq:18}
\begin{split}
&\dot{V_3}(t)\leq-\kappa_2\sum_{i=1}^NV_{ri}^2+s\sum_{i=1}^N\tilde{A_{Ti}}^2-\sum_{i=1}^N\tilde{A_{Ti}}V_{ri}(\sin\phi_i+\sigma_{2i})\\
&+\sum_{i=1}^N\tilde{A_{Ti}}V_{\lambda i}((1+\sum_{j=1}^N\frac{a_{ij}(R_iR_j\sin\theta_{ij})^2}{R_{i0}^2R_{j0}^2})\cos{\phi}_{i}-\sigma_{1i})\\
&\leq-\kappa_2\sum_{i=1}^NV_{ri}^2+s\sum_{i=1}^N\tilde{A_{Ti}}^2<0
\end{split}
\end{equation}
with $\sigma_{1i}=(1+\sum_{j=1}^N\frac{a_{ij}(R_iR_j\sin\theta_{ij})^2}{R_{i0}^2R_{j0}^2})\cos{\phi}_{i}$, $\sigma_{2i}=-\sin{\phi}_{i}, \kappa_2>0, i \in \textbf{\emph{V}}={1,2,...,N}$, and $s\leq0$.

Therefore, using the normal acceleration in formula~(\ref{eq:13}), the area of multiple attackers surrounding the target $S=\frac{1}{4}\sum_{i=1}^N\sum_{j=1}^Na_{ij}|R_iR_j\sin\theta_{ij}|$ and the angular velocity $\dot{\lambda}=\frac{V_\lambda}{R}$ of multiple attackers will converge to zero, and multiple attackers can attack the target at the same time at relatively uniform speed $V_r=-c$.
$\qed$
 }
 \end{pf}

   \begin{remark}
\label{rem4}
{\rm
The guidance laws ${A_{M\lambda}}$ and ${A_{Mr}}$ in equations~(\ref{eq:3},\ref{eq:13}) are distributed. Each attacker only needs the relative information of itself and its neighbors, and does not need to know the information of the whole attacker group. The main difference between the guidance laws in equation~(\ref{eq:3}) and that in equation~(\ref{eq:13}) is that the observation error term of acceleration is introduced, and the state convergence of the relative motion between the attackers and the target can be guaranteed under the condition of error disturbance.


It should be noted that the guidance laws of equation~(\ref{eq:3}) make the state $R$ converge to zero and the corresponding relative velocity $V_r$ converge to $-c$, that is, the remaining distance of relative motion decreases to zero continuously, and the relative velocity can be negative at this time. The guidance laws of equation~(\ref{eq:3}) make the state $R$ and the corresponding relative velocity $V_r$ converge to zero at the same time, that is, when the remaining distance of relative motion is reduced to zero, the corresponding relative velocity is likewise reduced to zero. In both cases, the guidance laws can accomplish simultaneous attack against a moving target by multiple attackers.
}
 \end{remark}

 \section{Simulation Results}
 In this section, to demonstrate the effectiveness of the proposed guidance laws, numerical simulations for a multi-UAV collaborative enclosure and simultaneous attack with known or unknown target acceleration are conducted. Initial parameters are listed in Table 1 and Table 2. Note that the superscript in Table 2 represents the data used in the corresponding example.

\subsection{Example 1: Cooperative attack with known target acceleration}

\begin{figure}[!hbt]
\centering
\includegraphics[width=3in]{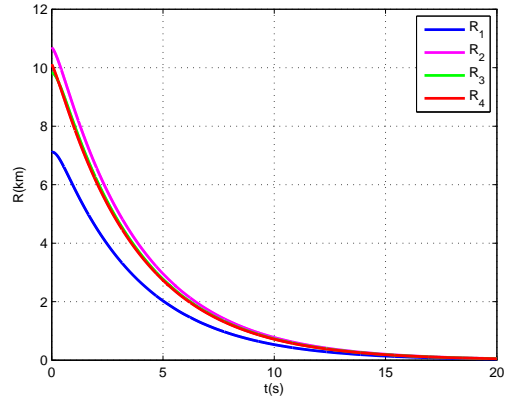}
\caption{Relative distances.\label{Fig4}}
  \label{Fig4}
\end{figure}

\begin{figure}[!hbt]
\centering
\includegraphics[width=3in]{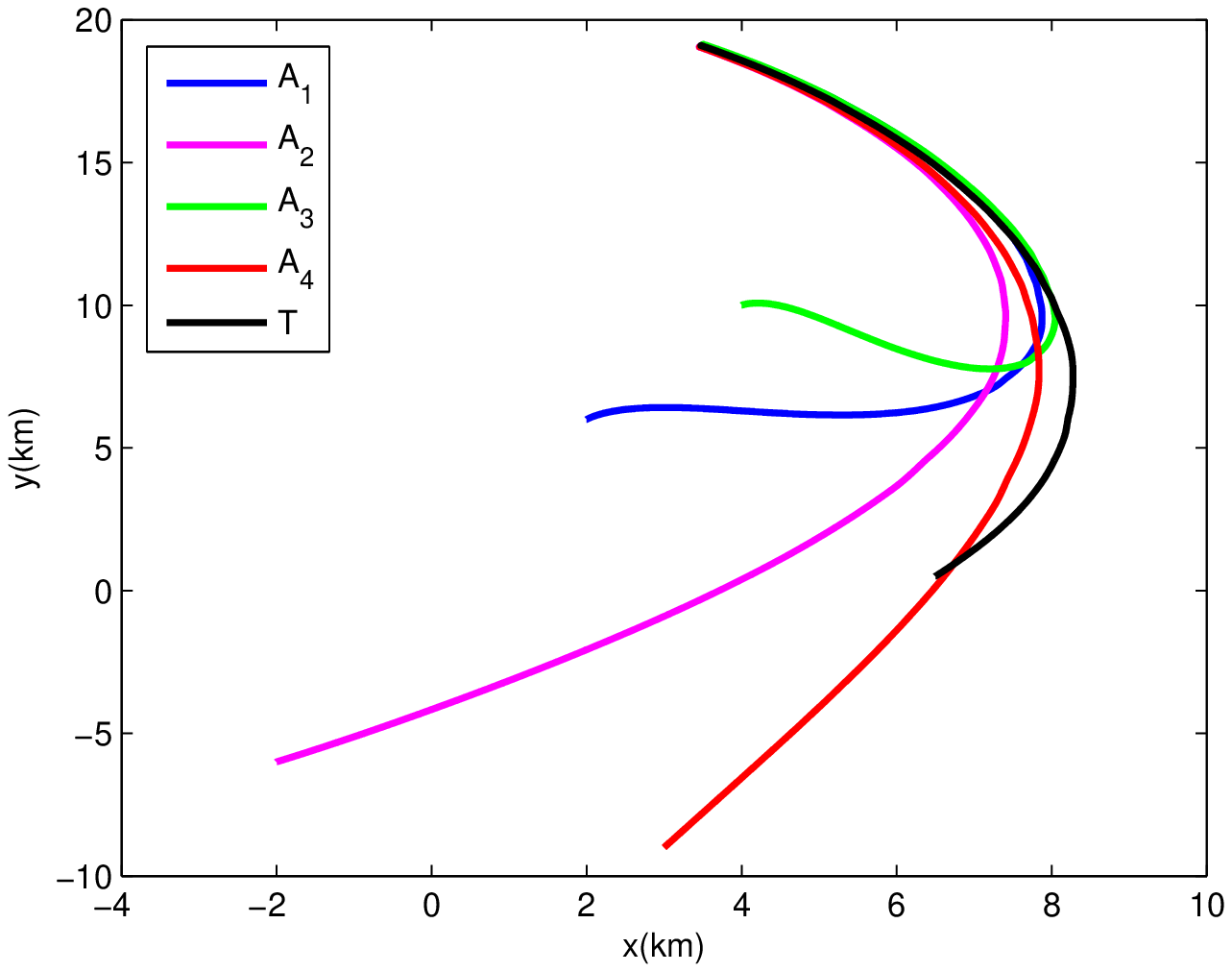}
\caption{Trajectories.\label{Fig5}}
  \label{Fig5}
\end{figure}

\begin{figure}[!hbt]
\centering
\includegraphics[width=3in]{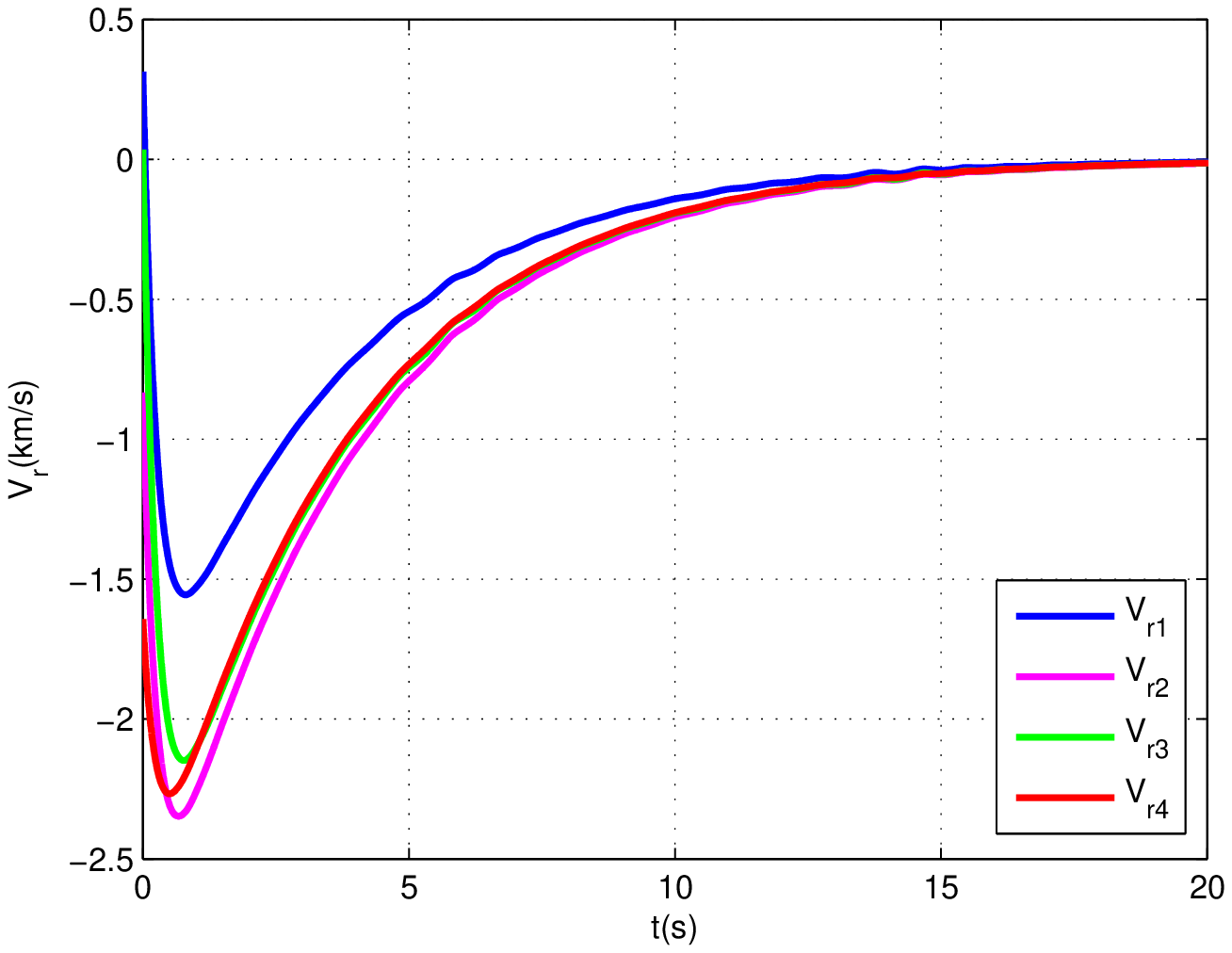}
\caption{Velocities ${V_r}$.\label{Fig6}}
  \label{Fig6}
\end{figure}

\begin{figure}[!hbt]
\centering
\includegraphics[width=3in]{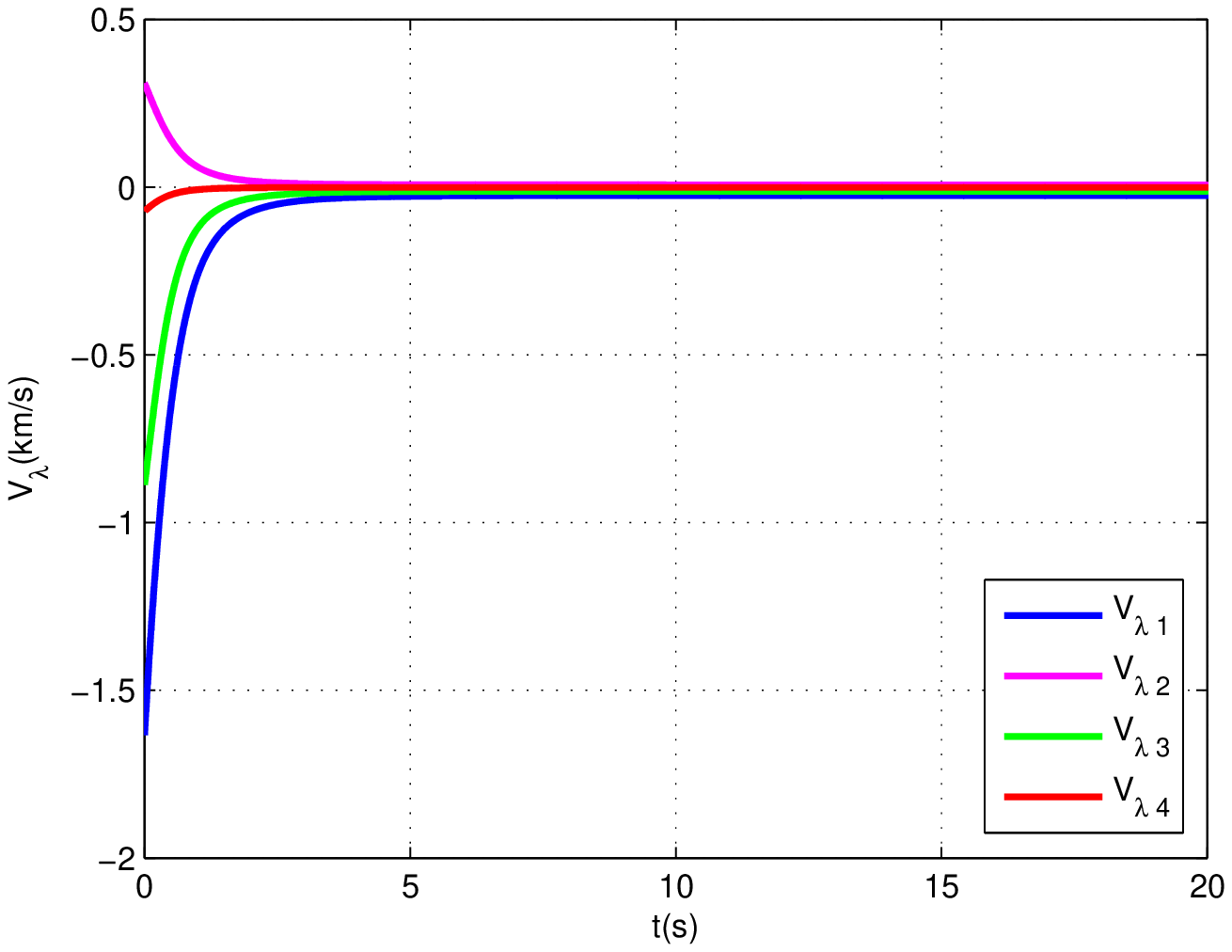}
\caption{Velocities ${V_\lambda}$.\label{Fig7}}
  \label{Fig7}
\end{figure}

\begin{figure}[!hbt]
\centering
\includegraphics[width=3in]{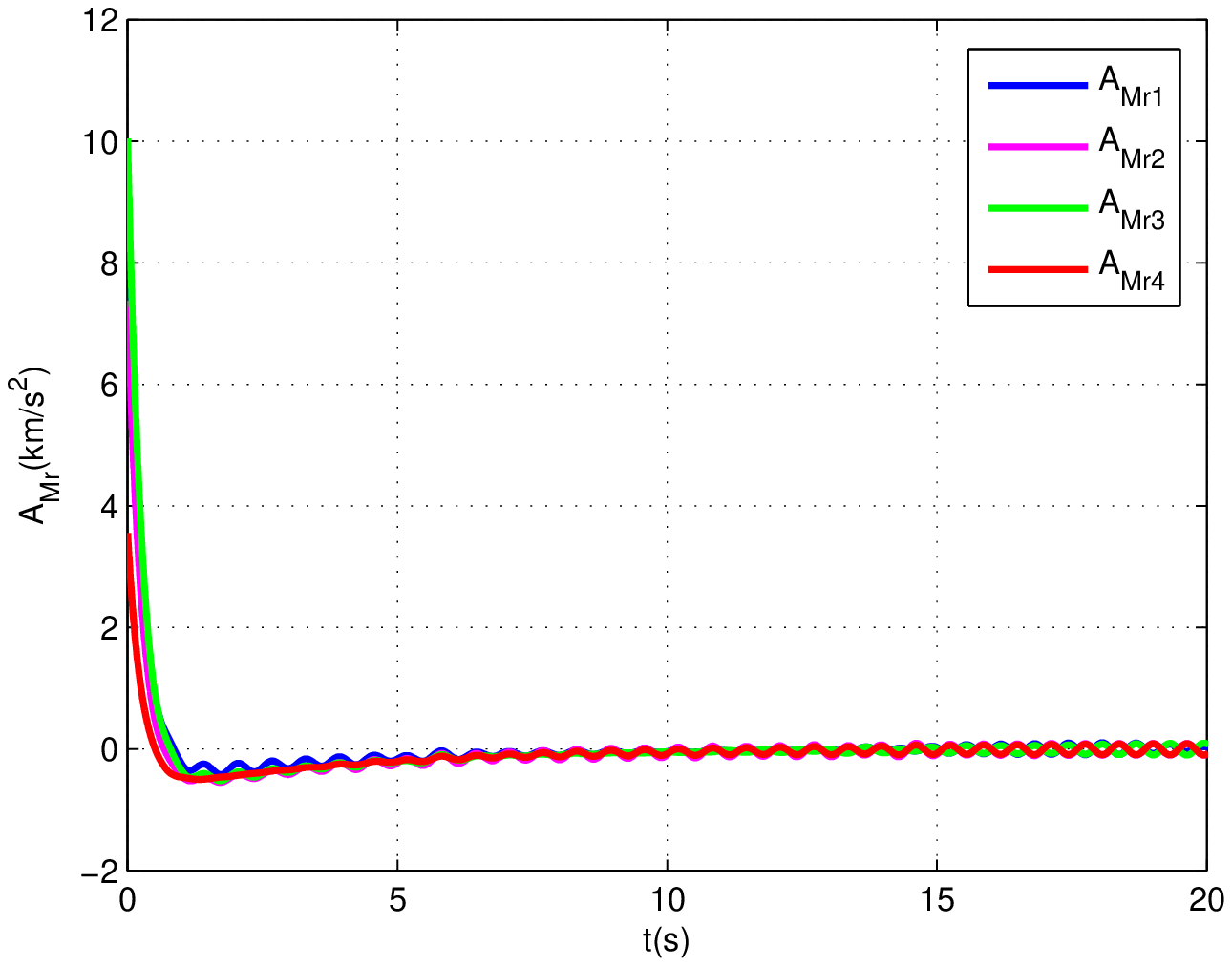}
\caption{Inputs ${A_{Mr}}$.\label{Fig8}}
  \label{Fig8}
\end{figure}

\begin{figure}[!hbt]
\centering
\includegraphics[width=3in]{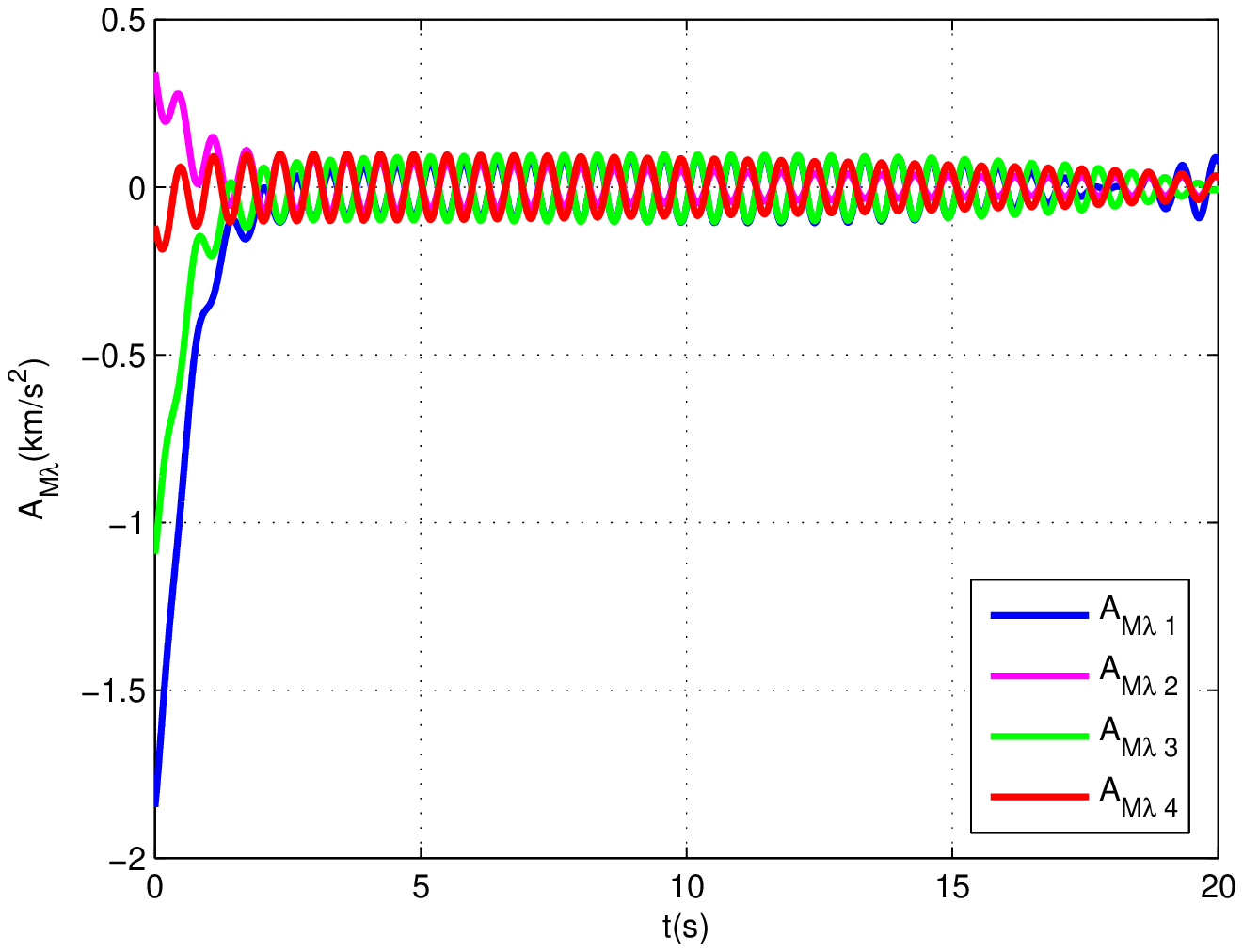}
\caption{Inputs ${A_{M\lambda}}$.\label{Fig9}}
  \label{Fig9}
\end{figure}

\begin{figure}[!hbt]
\centering
\includegraphics[width=3in]{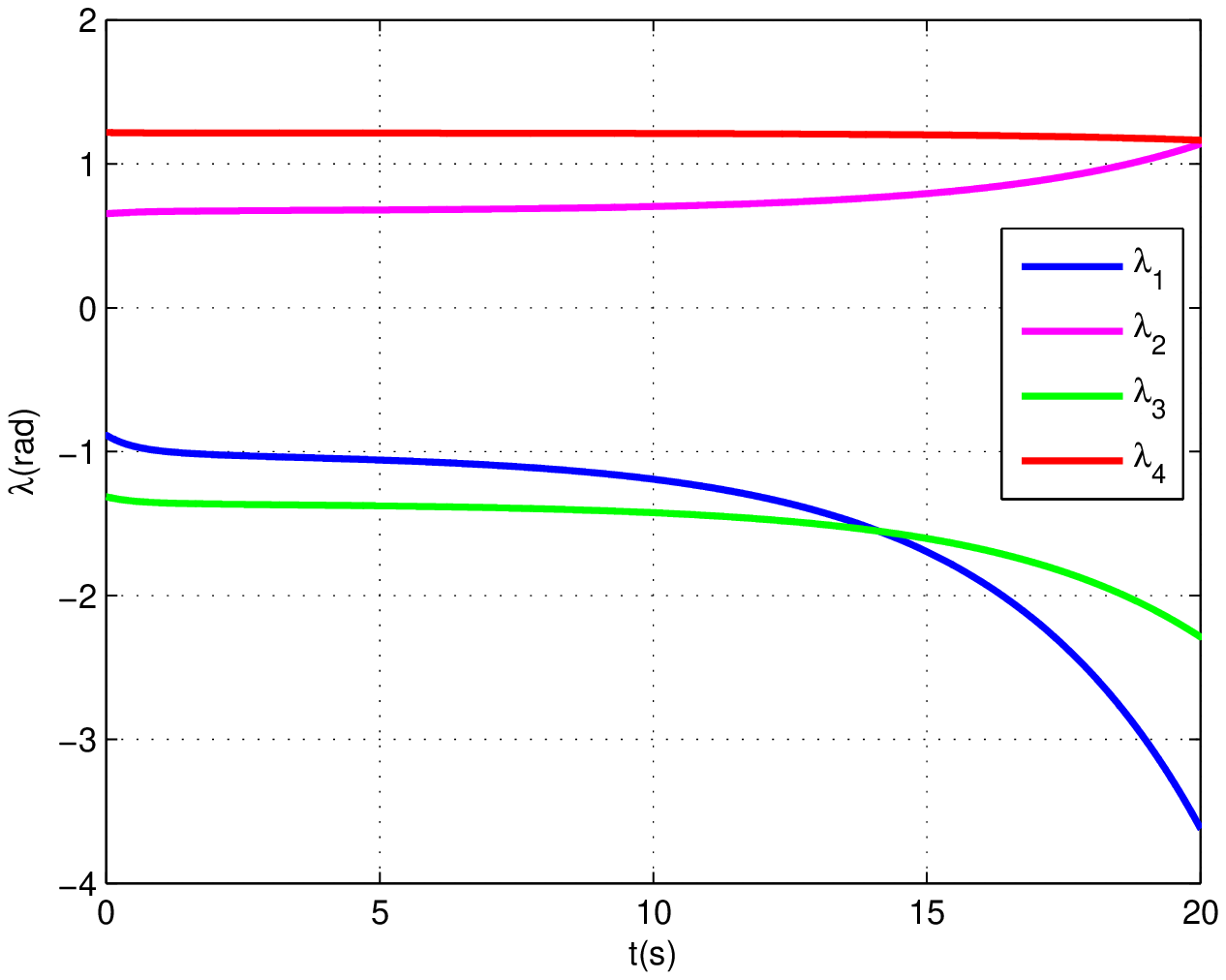}
\caption{Line of sight angle ${\lambda}$.\label{Fig10}}
  \label{Fig10}
\end{figure}

\begin{figure}[!hbt]
\centering
\includegraphics[width=3in]{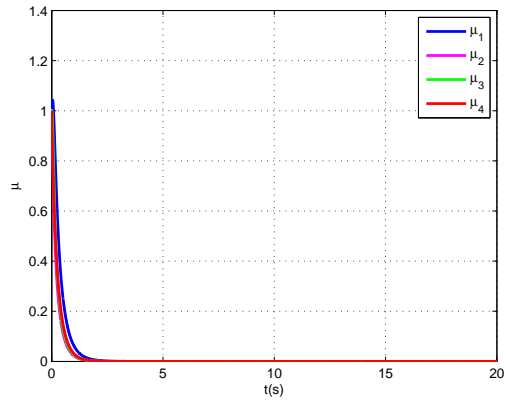}
\caption{Adaptive parameter $\mu$.\label{Fig11}}
  \label{Fig11}
\end{figure}

\begin{figure}[!hbt]
\centering
\includegraphics[width=3in]{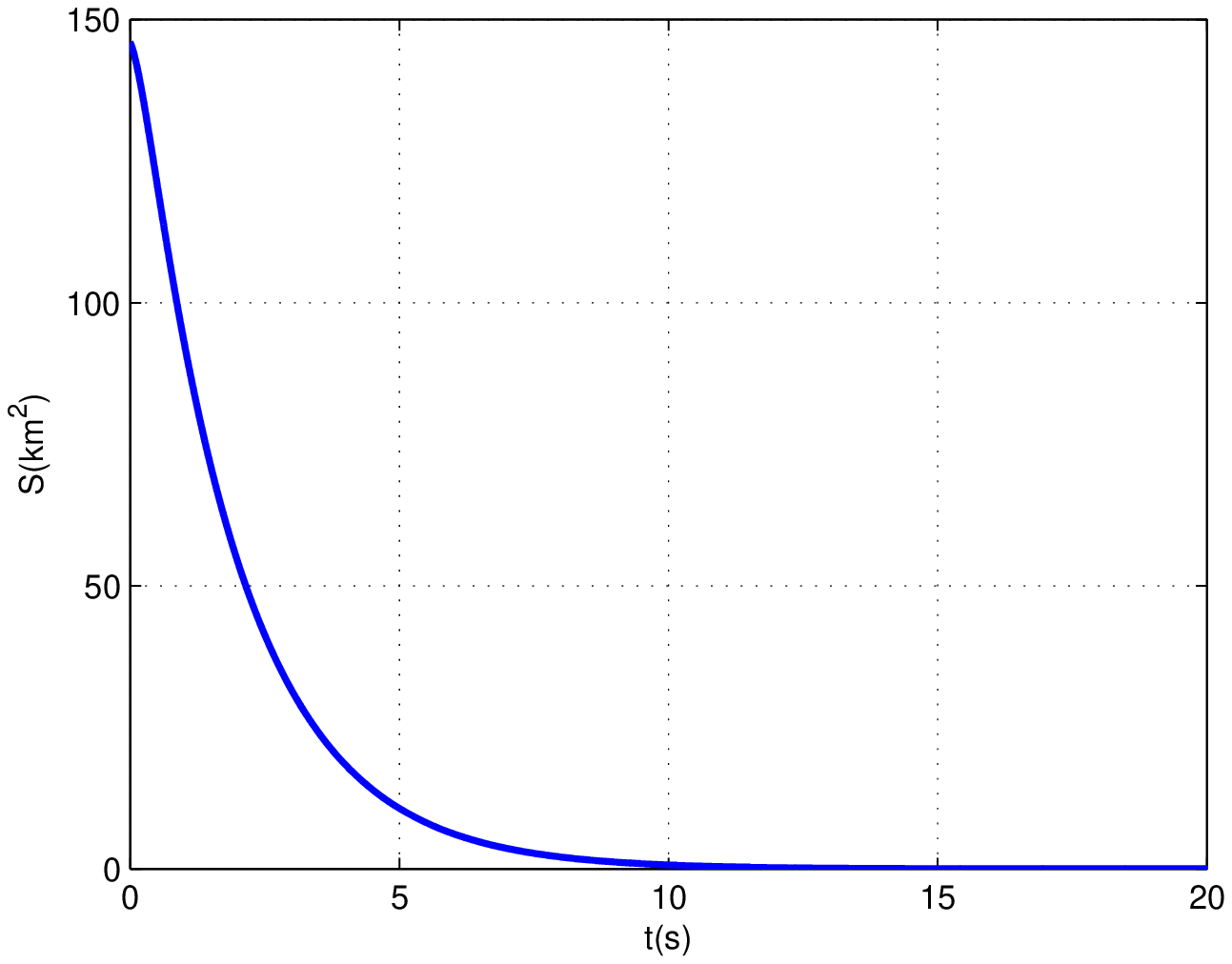}
\caption{Surrounding area $S$.\label{Fig12}}
  \label{Fig12}
\end{figure}


In this example, the acceleration component of the target is known, that is, the acceleration component along the direction of the target velocity is zero ($a_{tr}=0$ km/$s^2$), and the acceleration component perpendicular to the direction of the target velocity is time-varying ($a_{t\lambda}=0.1sin(10t)$ km/$s^2$). Then the acceleration components along and perpendicular to the LOS of the attackers are ${A_{Tr}}={a_{tr}}\cos{{\phi}}-a_{t\lambda}\sin{\phi}$ and ${A_{T\lambda}}=a_{tr}\sin{\phi}+a_{t\lambda}\cos{\phi}$. Attention should be paid to the fact that the accelerations of the target and the attacker are perpendicular to their respective velocity directions, which means that the speeds of the target and the attacker are constant and their direction are variable. The initial speeds of the target and the attacker are $V_i=0.7$ (km/s) and $V_T=1$ (km/s). In this example, four low-speed attackers attack a high-speed target at the same time. The guidance law in equation~(\ref{eq:3}) is adopted, in which the parameters are $\kappa_1=4$, $\kappa_2=4$, $\mu(t_0)=1$ and $c=0$.

Figures 3-12 depict the relative distance $R$, trajectory, relative velocity component along LOS $V_r$, relative velocity component perpendicular to LOS $V_\lambda$, input values ${A_{Tr}}$ and ${A_{T\lambda}}$, LOS angle $\lambda$, adaptive parameter $\mu$ and surrounding area $S$ of four low-speed attackers to strike a high-speed target at the same time. The simulation time is 20 seconds. It can be seen from the figures that the convergence of normal overload $\dot{V_\lambda}$ to zero makes the trajectory smooth, and the angular velocity of LOS $\dot{\lambda}$ is zero before the final strike, which means that multiple attackers can avoid internal collision in advance. The component of the attacker's acceleration along the LOS and the component of the attacker's acceleration perpendicular to the LOS have small chattering near zero. The reason is that the corresponding velocity component of acceleration control is zero, that is, when the velocity chatters near zero, the corresponding acceleration also has a small chattering near zero, and the acceleration components contain trigonometric functions ${A_{Mri}}=A_{Mi}\sin\xi_i$ and ${A_{M\lambda}}=-A_{Mi}\cos\xi_i$.

\subsection{Example 2: Cooperative attack with unknown target acceleration}

\begin{figure}[!hbt]
\centering
\includegraphics[width=3in]{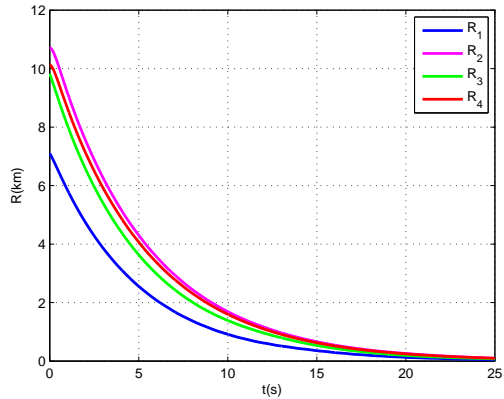}
\caption{Relative distances.\label{Fig13}}
  \label{Fig13}
\end{figure}

\begin{figure}[!hbt]
\centering
\includegraphics[width=3in]{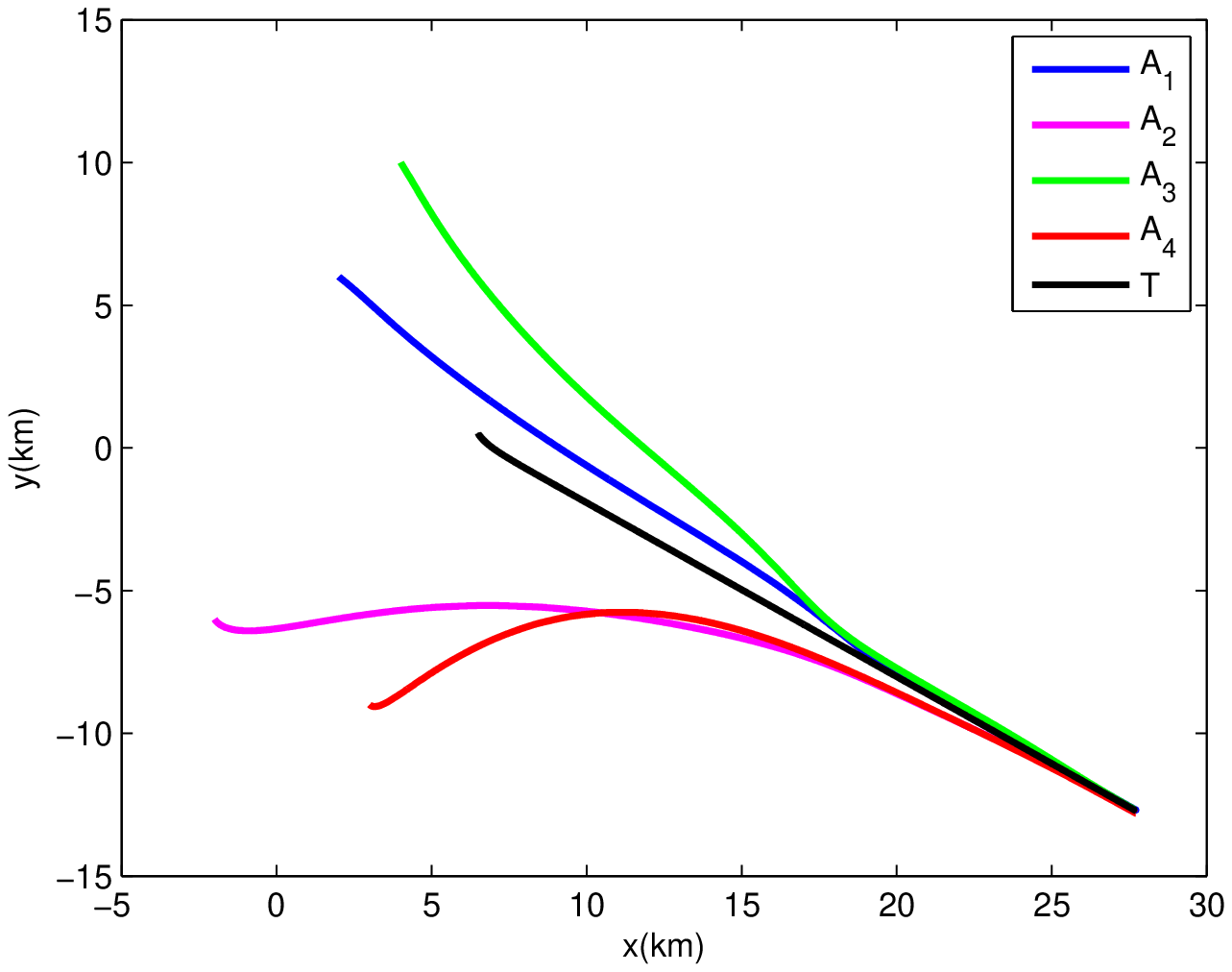}
\caption{Trajectories.\label{Fig14}}
  \label{Fig14}
\end{figure}

\begin{figure}[!hbt]
\centering
\includegraphics[width=3in]{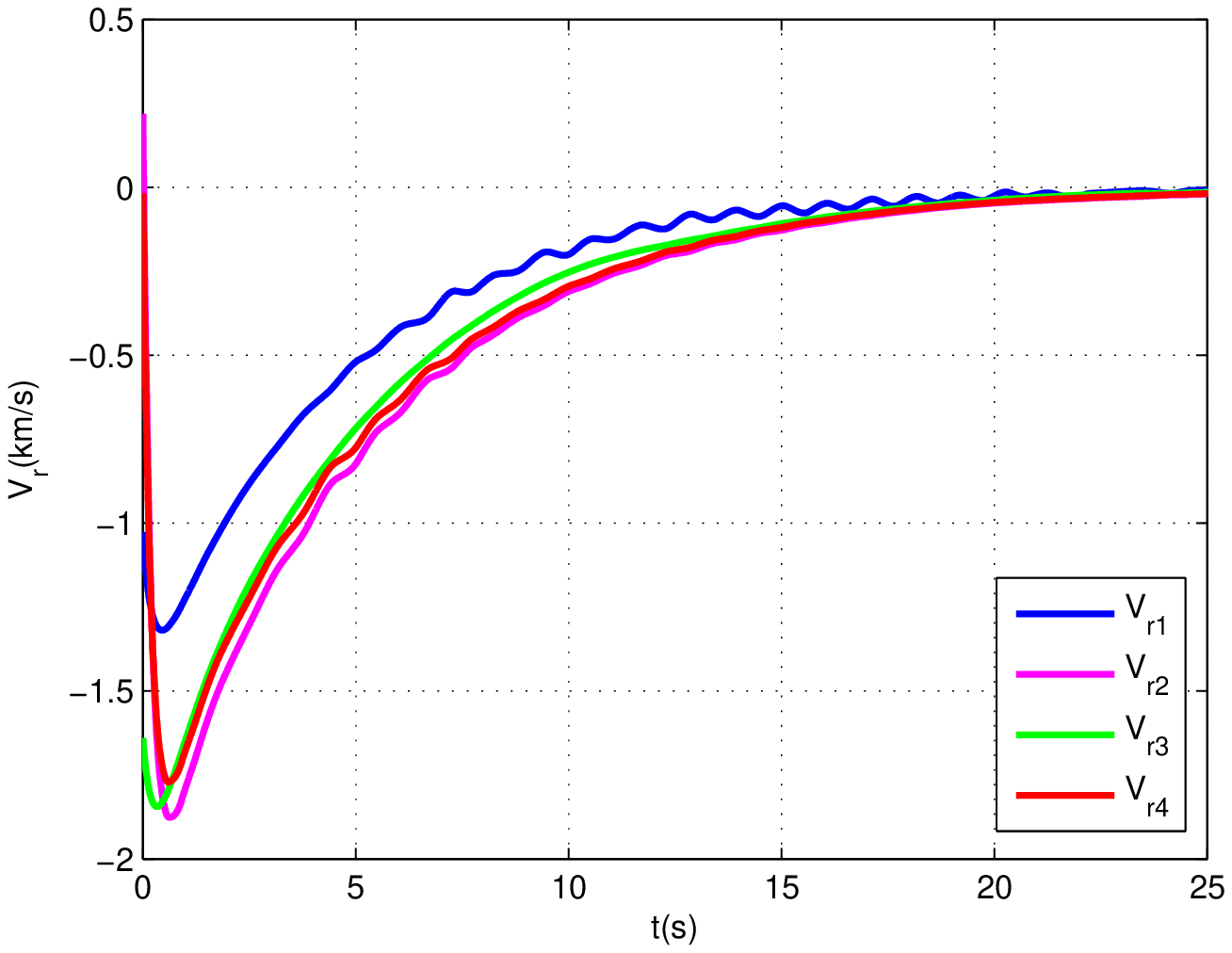}
\caption{Velocities ${V_r}$.\label{Fig15}}
  \label{Fig15}
\end{figure}

\begin{figure}[!hbt]
\centering
\includegraphics[width=3in]{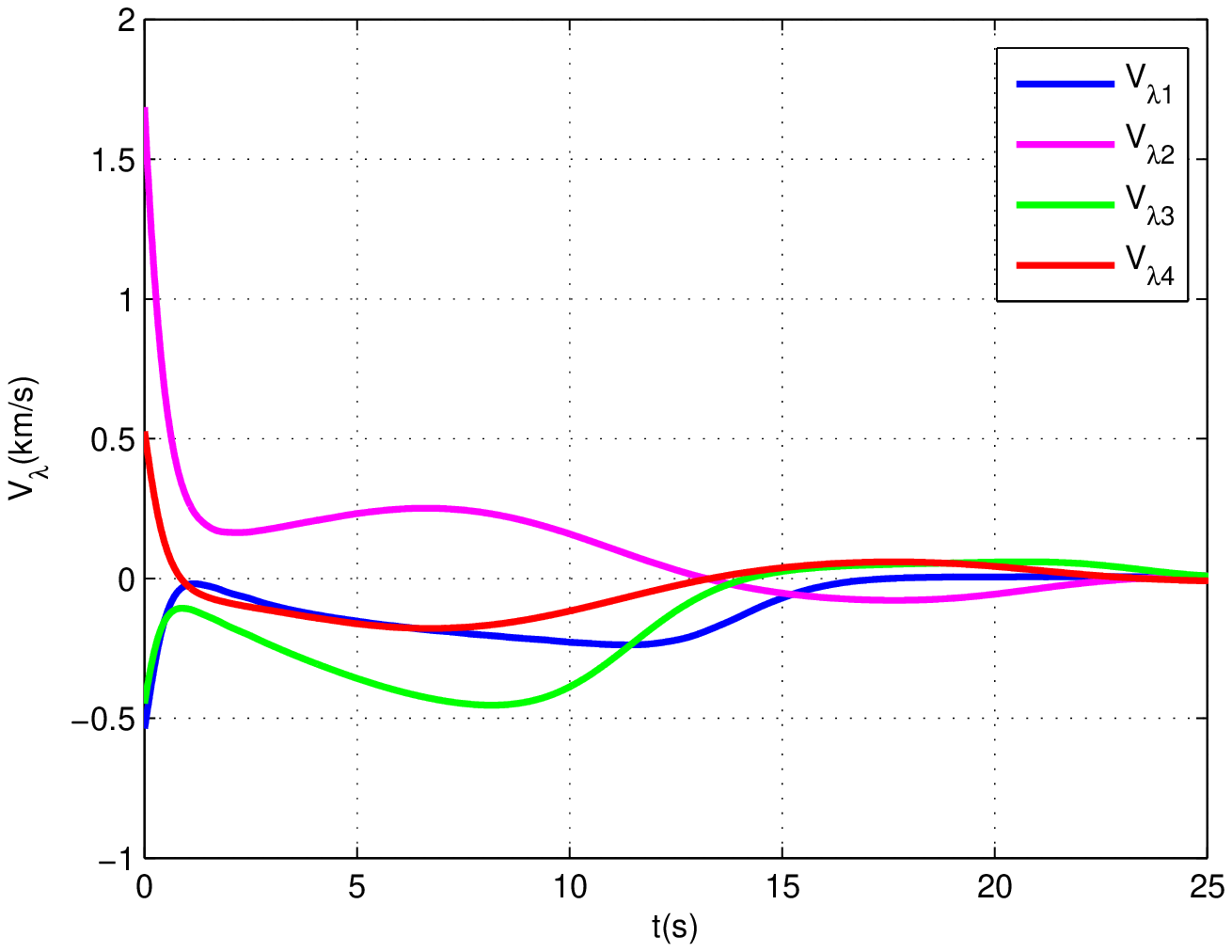}
\caption{Velocities ${V_\lambda}$.\label{Fig16}}
  \label{Fig16}
\end{figure}

\begin{figure}[!hbt]
\centering
\includegraphics[width=3in]{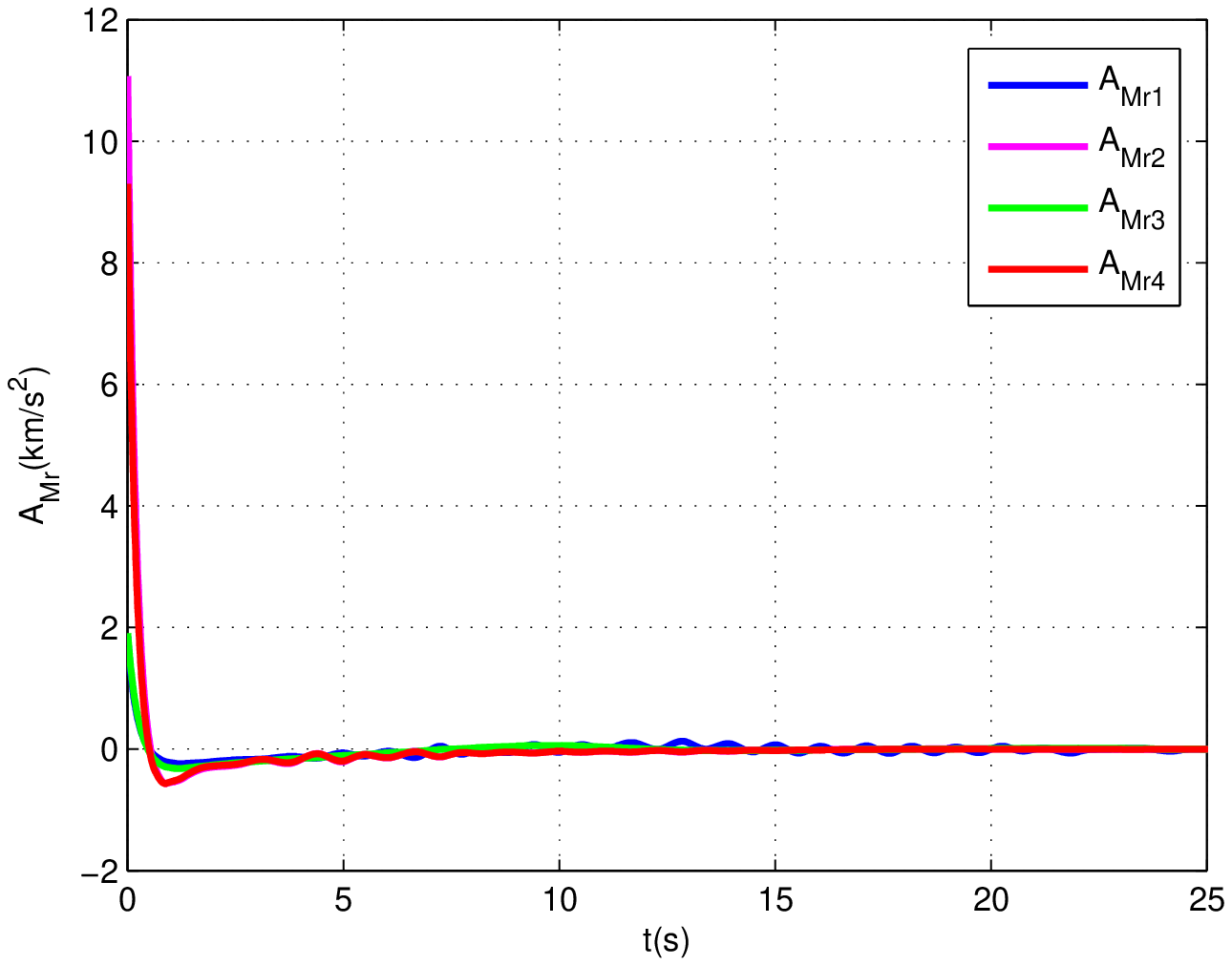}
\caption{Inputs ${A_{Mr}}$.\label{Fig17}}
  \label{Fig17}
\end{figure}

\begin{figure}[!hbt]
\centering
\includegraphics[width=3in]{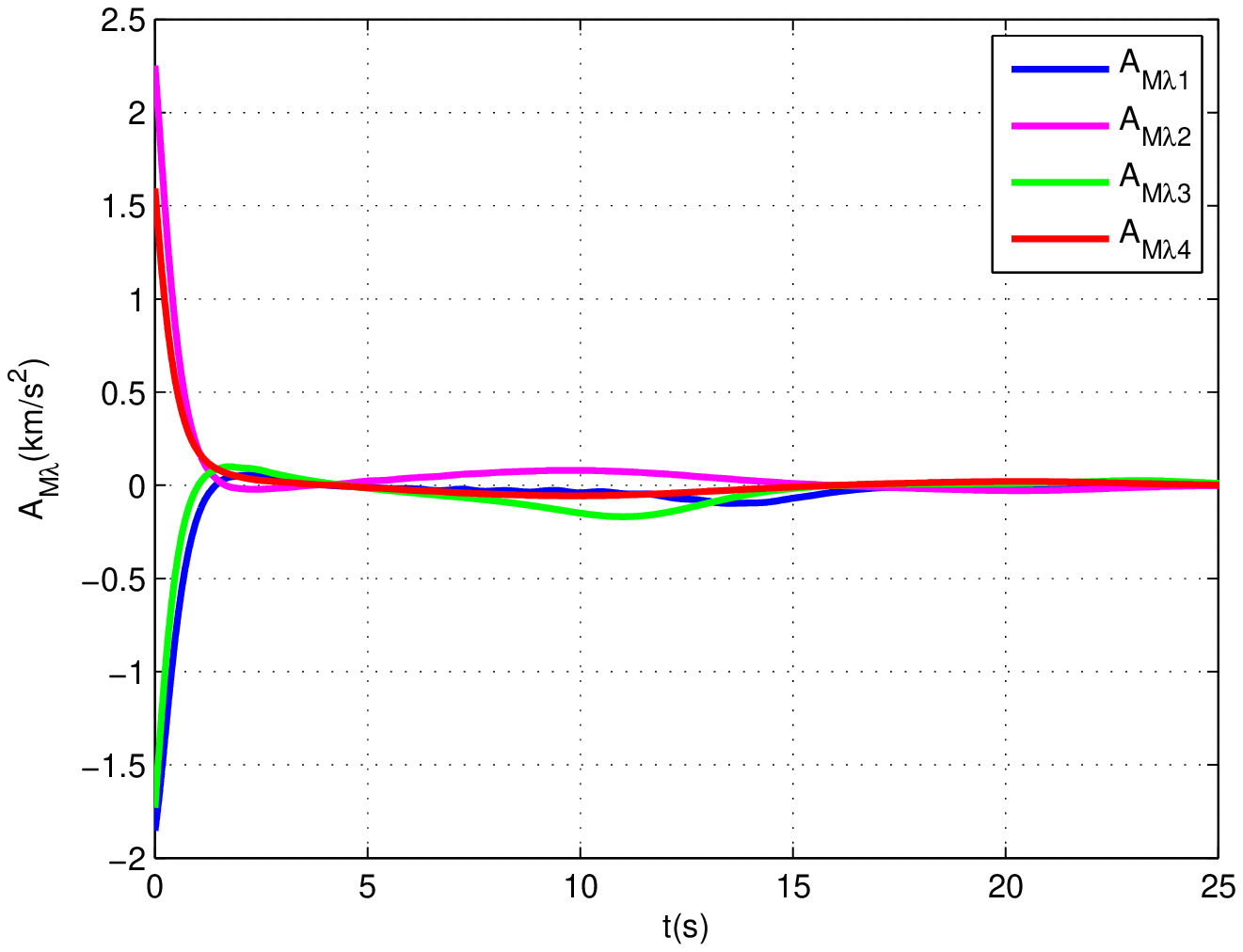}
\caption{Inputs ${A_{M\lambda}}$.\label{Fig18}}
  \label{Fig18}
\end{figure}

\begin{figure}[!hbt]
\centering
\includegraphics[width=3in]{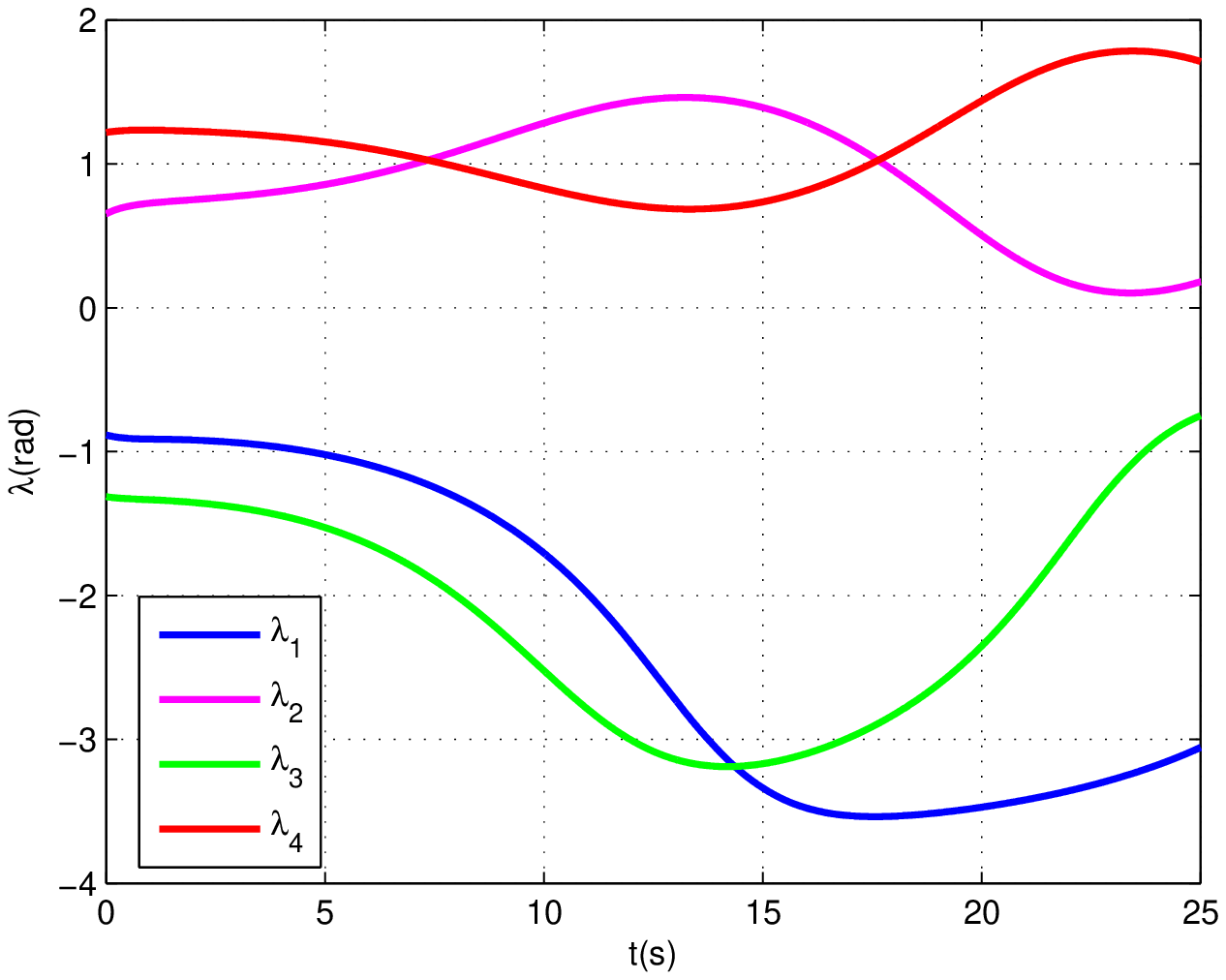}
\caption{Line of sight angle ${\lambda}$.\label{Fig19}}
  \label{Fig19}
\end{figure}

\begin{figure}[!hbt]
\centering
\includegraphics[width=3in]{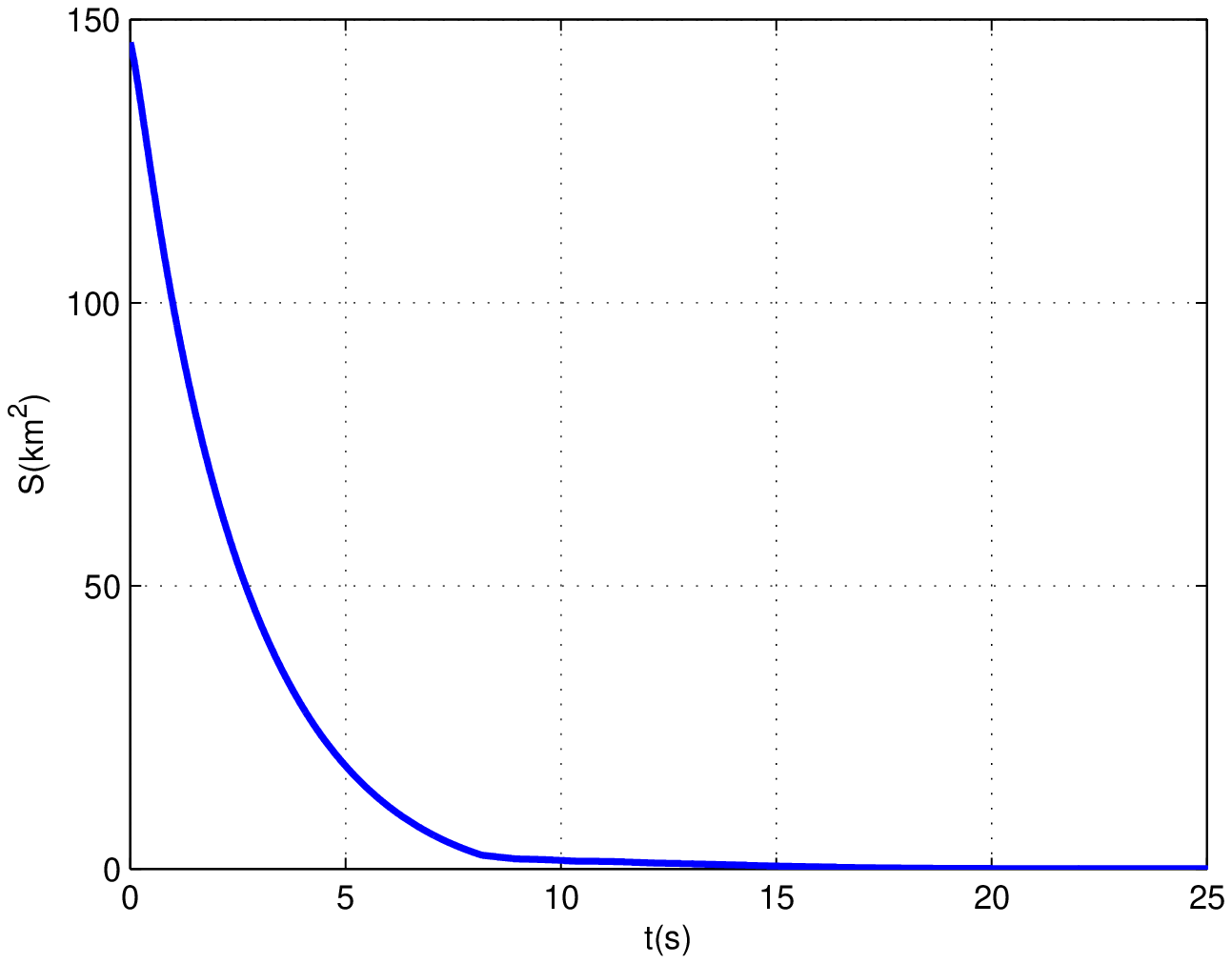}
\caption{Surrounding area $S$.\label{Fig20}}
  \label{Fig20}
\end{figure}


In this case, the target acceleration is unknown and its structure is in equation~(\ref{eq:11}) where $s=-2$. The initial conditions of this example are the same as those listed above. The guidance law in equation~(\ref{eq:13}) is adopted, in which the parameters are $\kappa_1=5$ and $\kappa_2=5$.


Figures 13-20 depict the relative distance $R$, trajectory, relative velocity component along LOS $V_r$, relative velocity component perpendicular to LOS $V_\lambda$, input values ${A_{Tr}}$ and ${A_{T\lambda}}$, LOS angle $\lambda$ and surrounding area $S$ of four low-speed attackers to strike a high-speed target at the same time. The simulation time is 25 seconds. Similar to the above, the relative distance and relative speed of the attacker-target are consistent, and the simultaneous attack task can be completed in a limited time.

\section{Conclusions}

In this paper, distributed guidance laws based on the enclosed area of a moving target by multiple low-speed attackers are designed to coordinate multiple attackers to surround or simultaneously attack a high-speed target, whose acceleration can be observed. As long as at least one attacker can observe the information of the target, the other attackers can obtain the information of the relative movement of the attacker-target indirectly from the communication network that contains a spanning tree, without all attackers observing the target. The novel guidance laws can converge the normal overload of attacker-target relative motion to zero, which ensures the smoothness of the attacker's trajectory. It can also avoid the two major problems in designing of guidance laws for multiple attackers attacking a moving target at the same time: the difficulty in calculation of remaining time and the avoidance of collision. In this paper, only the point mass model of multi-UAV system is studied, and its attitude control will be discussed in the future work.

\begin{table}
 \caption{\label{tab:table1}Simulation parameters of four attackers }
 \begin{ruledtabular}
 \begin{tabular}{llllll}
  $Parameters$ & ${Attacker}_1$ & ${ Attacker}_2$ & ${ Attacker}_3$ & ${ Attacker}_4$ \\ \hline
   $\lambda,rad$    &-0.8851   & 0.6528  & -1.3135  &  1.2178 \\
   $\gamma,rad$    & 0.6283 &-1.0472 &-1.0472 &1.5708  \\
   $R,km$ & 7.1063 & 10.7005& 9.8234 &10.1242 \\
  \end{tabular}
  \end{ruledtabular}
\end{table}

\begin{table}
 \caption{\label{tab:table2}Simulation parameters of the target }
 \begin{ruledtabular}
 \begin{tabular}{llll}
  $Target_x,km$ & $Target_y,km$ & $\sigma_T^1,rad$ & $\sigma_T^2,rad$ \\ \hline
   6.5000 & 0.5000 & 1.0472 & -1.0472 \\
  \end{tabular}
  \end{ruledtabular}
\end{table}

\section*{Acknowledgments}
This research was supported by the National Natural Science Foundation of China under Grant No. 11332001 and No.61773024, Innovation Research Project Fund 17-163-11-ZT-003-018-01, and Joint Fund of the Ministry of Education of China 6141A020223.


\section*{References}
\begin{thebibliography}{0}

\bibitem{1}
Chen, F., and Ren, W., ``A connection between dynamic region-following formation control and distributed average tracking,'' \textit{IEEE Trans. on Cybernetics}, Vol.48, No.6, 2018, pp. 1760--1772.
DOI: 10.1109/TCYB.2017.2714688

\bibitem{2}
Dong, X., and Hu, G., ``Time-varying formation control for general linear multi-agent systems with switching directed topologies,'' \textit{Automatica}, Vol.73, No.C, 2016, pp. 7--55.
DOI: 10.1016/j.automatica.2016.06.024

\bibitem{3}
Wang, B., Wang, J., Zhang, B., and Li, X., ``Global cooperative control framework for multiagent systems subject to actuator saturation with industrial applications,'' \textit{IEEE Trans. on Systems~ Man and Cybernetics, Systems}, Vol.47, No.7, 2017, pp. 1270--1283.
DOI: 10.1109/tsmc.2016.2573584

\bibitem{4}
Zhou, J., and Yang, J., ``Distributed guidance law design for cooperative simultaneous attacks with multiple missiles,'' \textit{Journal of Guidance Control and Dynamics}, Vol.39, No.10, 2016, pp. 1--9.
DOI: 10.2514/1.g001609 


\bibitem{5}
Hou, D., Wang, Q., Sun, X., and Dong, C., ``Finite-time cooperative guidance laws for multiple missiles with acceleration saturation constraints,'' \textit{Control Theory and Applications Iet}, Vol.9, No.10, 2015, pp. 1525--1535.
DOI: 10.1049/iet-cta.2014.0443 


\bibitem{6}
Wang, X., Zhang, Y., and Wu, H., ``Distributed cooperative guidance of multiple anti-ship missiles with arbitrary impact angle constraint,'' \textit{Aerospace Science and Technology}, Vol.46, 2015, pp. 299--311.
DOI: 10.1016/j.ast.2015.08.002


\bibitem{7}
Bing, Z., Zaini, A. H. B., and Xie, L., ``Distributed guidance for interception by using multiple rotary-wing unmanned aerial vehicles,'' \textit{IEEE Transactions on Industrial Electronics}, Vol.64, No.7, 2017, pp. 5648--5656.
DOI: 10.1109/AIM.2016.7576906 

\bibitem{8}
Zhou, J., Yang, J., and Li, Z., ``Simultaneous attack of a stationary target using multiple missiles, a consensus-based approach,'' \textit{Science China Information Sciences},  Vol.60, No.7, 2017, pp. 67--80.
DOI: 10.1007/s11432-016-9089-7

\bibitem{9}
Jeon, I. S., Lee, J. I., and Tahk, M. J., ``Homing Guidance Law for Cooperative Attack of Multiple Missiles,'' \textit{ Journal of Guidance Control and Dynamics}, Vol.33, No.1, 2010, pp. 275--280.
DOI: 10.2514/1.40136

\bibitem{10}
Wang, X., Zheng, Y., and Lin, H., ``Integrated guidance and control law for cooperative attack of multiple missiles,'' \textit{Aerospace Science and Technology}, Vol.42, 2015, pp. 1--11.
DOI: 10.1016/j.ast.2014.11.018 

\bibitem{11}
Sun, X., and Xia, Y., ``Optimal guidance law for cooperative attack of multiple missiles based on optimal control theory,'' \textit{International Journal of Control}, Vol.85, No.8, 2012, pp. 1063--1070.
DOI: 10.1080/00207179.2012.675519

\bibitem{12}
Kang, S., Wang, J., Li, G., Shan, J., and Petersen, I.R., ``Optimal cooperative guidance law for salvo attack: an MPC-based consensus perspective,'' \textit{IEEE Transactions on Aerospace and Electronic Systems}, Vol.54, No.5, 2018, pp. 2397--2410.
DOI: 10.1109/TAES.2018.2816880


\bibitem{13}
Kim, H. G., Cho, D., and Kim, H. J., ``Sliding mode guidance law for impact time control without explicit time-to-go estimation,'' \textit{IEEE Transactions on Aerospace and Electronic Systems}, Vol.55, No.1, 2019, pp. 236--250.
DOI: 10.1109/TAES.2018.2850208

\bibitem{14}
Zhou, J., Lv, Y., Li, Z., and Yang, J., ``Cooperative guidance law design for simultaneous attack with multiple missiles against a maneuvering target,'' \textit{Journal of Systems Science and Complexity},  Vol. 31, No. 1, 2018, pp. 287--301.
DOI: 10.1007/s11424-018-6317-7

\bibitem{15}
Hull, D. G., Radke, J. J., and Mack, R. E., ``Time-to-go prediction for homing missiles based on minimum-time intercepts,'' \textit{Journal of Guidance Control and Dynamics},  Vol.14, No.5, 2012, pp. 865--871.
DOI: 10.2514/3.20725

\bibitem{16}
Ding, S., and Zheng, W., ``Nonsingular terminal sliding mode control of nonlinear second-order systems with input saturation,'' \textit{ International Journal of Robust and Nonlinear Control}, Vol. 26, 2016, pp. 1857--1872.
DOI: 10.1002/rnc.3381


\bibitem{17}
Lee, C. H., Kim, T. H., and Tahk, M. J., ``Interception angle control guidance using proportional navigation with error feedback,'' \textit{Journal of Guidance Control and Dynamics}, Vol. 36, No. 5, 2013, pp. 1556--1561.
DOI: 10.2514/1.58454

\bibitem{18}
Lee, Y. I., Kim, S. H., and Tahk, M. J., ``Optimality of linear time-varying guidance for impact angle control,'' \textit{IEEE Transactions on Aerospace and Electronic Systems}, Vol. 48, No. 4, 2012, pp. 2802--2817.
DOI: 10.1109/TAES.2012.6324662

\bibitem{19}
Cho, N., and Kim, Y., ``Modified pure proportional navigation guidance law for impact time control,'' \textit{Journal of Guidance Control and Dynamics},  Vol. 39, No. 4, 2016, pp. 1--21.
DOI: 10.2514/1.G001618


\bibitem{20}
Alkaher, D., Moshaiov, A., and Or,Y., ``Guidance laws based on optimal feedback linearization pseudocontrol with time-to-go estimation,'' \textit{Journal of Guidance Control and Dynamics}, Vol.37, No.4, 2014, pp. 1298--1305.
DOI: 10.2514/1.g000205

\bibitem{21}
Lee, C. H., Kim, T. H., and Tahk, M. J., ``Effects of time-to-go errors on performance of optimal guidance laws,'' \textit{IEEE Transactions on Aerospace and Electronic Systems}, Vol.51, No.4, 2015, pp. 3270--3281.
DOI: 10.1109/taes.2015.150163


\end{thebibliography}
\end{document}